\documentclass{emulateapj}

\newcommand\kms{\ifmmode{\rm km\thinspace s^{-1}}\else km\thinspace s$^{-1}$\fi}
\newcommand\hip{{\it Hipparcos\/}}
\newcommand\pht{\phantom{22}}
\newcommand\an{AN\ }

\shortauthors{Torres et al.}
\shorttitle{LV~Her}

\begin{document}

\journalinfo{Accepted for publication in The Astronomical Journal}

\title{Absolute properties of the highly eccentric eclipsing binary
star LV~Herculis}

\author{
Guillermo Torres\altaffilmark{1},
Claud H.\ Sandberg Lacy\altaffilmark{2},
Antonio Claret\altaffilmark{3}
}

\altaffiltext{1}{Harvard-Smithsonian Center for Astrophysics, 60
Garden St., Cambridge, MA 02138, USA; e-mail: gtorres@cfa.harvard.edu}

\altaffiltext{2}{Department of Physics, University of Arkansas,
Fayetteville, AR 72701, USA; e-mail: clacy@uark.edu}

\altaffiltext{3}{Instituto de Astrof\'\i sica de Andaluc\'\i a, CSIC,
Apartado 3004, 18080 Granada, Spain; e-mail: claret@iaa.es}

\begin{abstract} 

We report extensive spectroscopic and differential $V$-band
photometric observations of the 18.4-day detached double-lined
eclipsing binary LV~Her (\ion{F9}{5}), which has the highest
eccentricity ($e \simeq 0.613$) among the systems with well-measured
properties. We determine the absolute masses and radii of the
components to be $M_1 = 1.193 \pm 0.010~M_{\sun}$, $M_2 = 1.1698 \pm
0.0081~M_{\sun}$, $R_1 = 1.358 \pm 0.012~R_{\sun}$, and $R_2 = 1.313
\pm 0.011~R_{\sun}$, with fractional errors of 0.9\% or better. The
effective temperatures are $6060 \pm 150$~K and $6030 \pm 150$~K,
respectively, and the overall metallicity is estimated to be [m/H] $=
+0.08 \pm 0.21$. A comparison with current stellar evolution models
for this composition indicates an excellent fit for an age between 3.8
and 4.2~Gyr, with both stars being near the middle of their
main-sequence lifetimes.  Full integration of the equations for tidal
evolution is consistent with the high eccentricity, and suggests the
stars' spin axes are aligned with the orbital axis, and that their
rotations should be pseudo-synchronized. The latter prediction is not
quite in agreement with the measured projected rotational velocities.

\end{abstract}

\keywords{
binaries: eclipsing --- 
binaries: spectroscopic --- 
stars: evolution --- 
stars: fundamental parameters --- 
stars: individual (LV~Her)
}

\section{Introduction}
\label{sec:introduction}

The photometric variability of the eclipsing binary star LV~Her
(TYC~2076-1042-1; $\alpha = 17^{\rm h}\,35^{\rm m}\,32\fs40$, $\delta
= +23\arcdeg\,10\arcmin\,30\farcs6$, J2000.0; SpT F9, $V = 11.02$) was
announced by \cite{Hoffmeister:35}. Its period was estimated by
\cite{Zessewitsch:44} as 2.634 days, by \cite{Zessewitsch:54} as
5.2674 days, by \cite{Popper:96} as 9.218 days, by \cite{Torres:00} as
18.1312 days, and by \cite{Torres:01} as 18.4359350 days. All
photometric and spectroscopic work performed since this last study has
confirmed that it is basically correct. The prior erroneous period
values resulted from the very long orbital period, the similar eclipse
depths (0.68 mag and 0.66 mag in $V$; see below), the extremely high
eccentricity of the orbit (0.613), and the resultant shift of
secondary eclipse to a phase of 0.86. This sort of erroneous period
estimation is not unusual in this kind of situation \citep{Lacy:04a,
Lacy:04b, Lacyetal:06}.

Largely because of this uncertainty, no determination of the absolute
dimensions of the system has been made until now. In addition to an
accurate determination of the eclipse ephemeris in
\S\thinspace\ref{sec:ephemeris}, this work presents new high-quality
photometric and spectroscopic observations
(\S\thinspace\ref{sec:photometry} and
\S\thinspace\ref{sec:spectroscopy}) that yield masses and radii on a
par with the best determinations for eclipsing binaries to date
(\S\thinspace\ref{sec:dimensions}). In \S\thinspace\ref{sec:evolution}
we compare these determinations with models of stellar structure and
stellar evolution. The high eccentricity and long period of LV~Her
make it an interesting case for comparison with tidal theory, which we
present in \S\thinspace\ref{sec:tidal}. We discuss our results in
\S\thinspace\ref{sec:discussion}.

\section{Observations and reductions}

\subsection{Differential and absolute photometry}
\label{sec:photometry}

Our photometric work on LV~Her began in 2001 and is based on
observations obtained with two robotic instruments: the URSA telescope
at Kimpel Observatory \citep{Lacy:01} on the campus of the University
of Arkansas, and a robotic telescope at the NF Observatory
\citep[NFO;][]{Grauer:08} near Silver City, NM.

\begin{figure}[b] 
\epsscale{1.1} 
\plotone{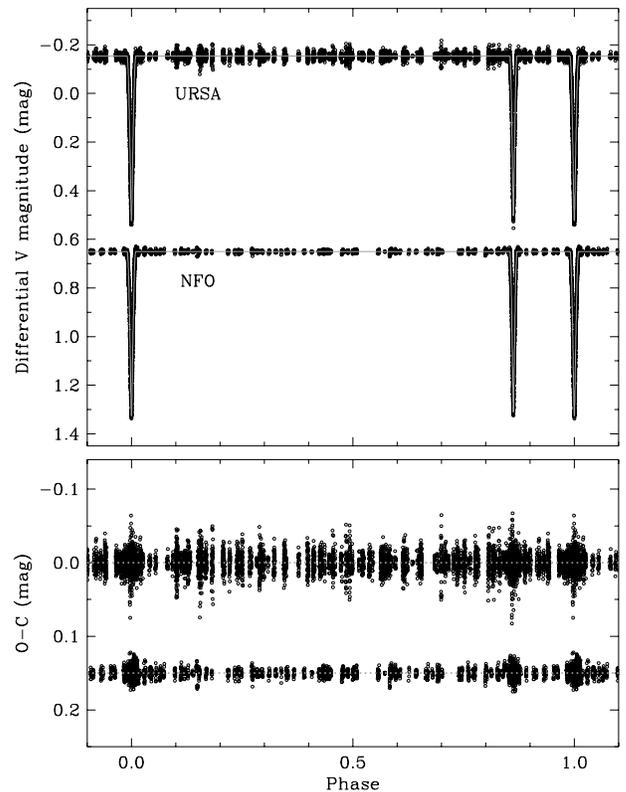}
%
\figcaption[]{URSA and NFO $V$-band photometry for LV~Her, shown with
our best fit model described in \S\thinspace\ref{sec:lcfits}. Residuals are
shown at the bottom, with NFO displaced for clarity.\label{fig:LCall}}
\end{figure}

\begin{figure}
\epsscale{1.1} 
\plotone{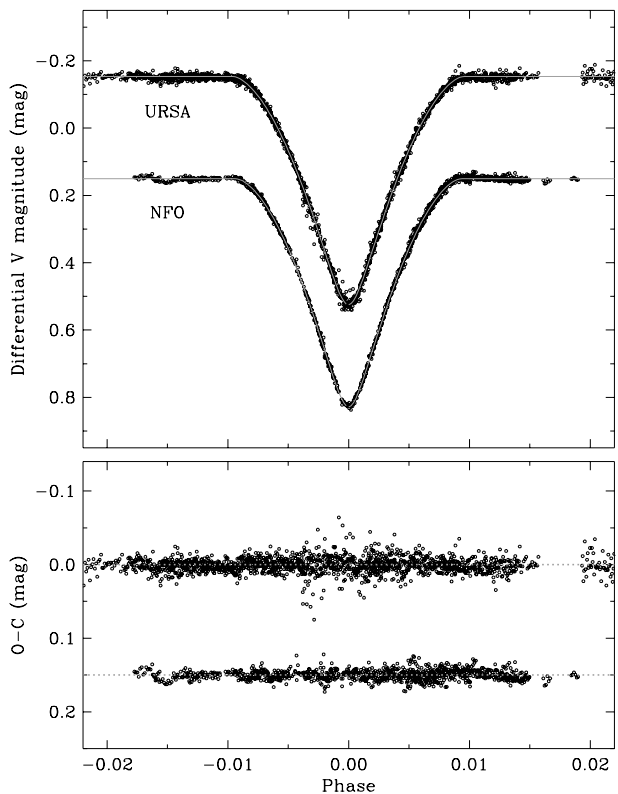}
%
\figcaption[]{Enlarged view of the URSA and NFO $V$-band light curves
for LV~Her near the primary eclipse, shown with our best fit model
described in \S\thinspace\ref{sec:lcfits}. Residuals are shown at the
bottom, with NFO displaced for clarity.\label{fig:LCprim}}
\end{figure}

\begin{figure} 
\epsscale{1.1} 
\plotone{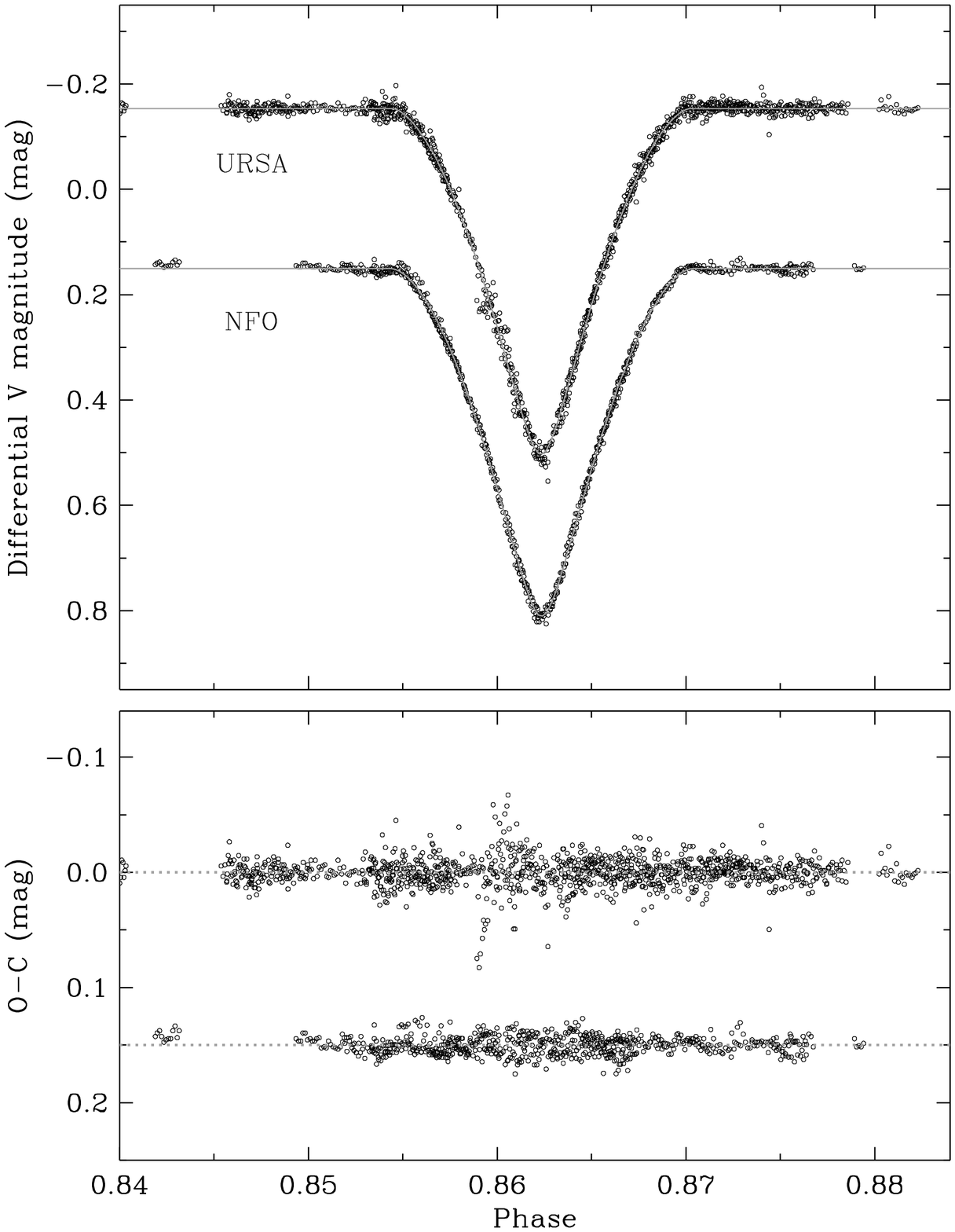}
%
\figcaption[]{Enlarged view of the URSA and NFO $V$-band light curves
for LV~Her near the secondary eclipse, shown with our best fit model
described in \S\thinspace\ref{sec:lcfits}. Residuals are shown at the
bottom, with NFO displaced for clarity.\label{fig:LCsec}}
\end{figure}

Kimpel Observatory (ursa.uark.edu) consists of a Meade 10-inch f/6.3
LX-200 telescope with a Santa Barbara Instruments Group ST8 CCD camera
(binned $2\times2$ to produce $765 \times 510$ pixel images with 2.3
arcsec square pixels) inside a Technical Innovations Robo-Dome, and
controlled automatically by an Apple Macintosh G4 computer.  The
observatory is located on top of Kimpel Hall on the Fayetteville
campus of the University of Arkansas, with the control room directly
beneath the observatory inside the building.  Exposures of 60 or 120
seconds made through a Bessell $V$ filter (2.0 mm of GG~495 and 3.0 mm
of BG~39) were read out and downloaded by ImageGrabber (camera control
software written by J.\ Sabby) to the control computer over a 30
second interval, then the next exposure was begun.  The observing
cadence was therefore about 90 to 150 sec per observation.  The
variable star would sometimes be monitored continuously for 4--6
hours.  LV~Her was observed by URSA on 156 nights during parts of nine
observing seasons from 2001 February 18 to 2009 April 8. The total
number of URSA observations is 6690.  The frames were analyzed by a
virtual measuring engine application written by Lacy that flat-fielded
the images, automatically located the variable and comparison stars in
the image, measured their brightnesses, subtracted the corresponding
sky brightness, and corrected for the differences in airmass between
the stars.  Extinction coefficients were determined nightly from the
comparison star measurements.  They averaged 0.25 mag/airmass.  LV~Her
is TYC~2076-1042-1.  The comparison stars were TYC~2076-0580-1
\citep[comp, $V_T = 11.12$, as listed in the Tycho-2
Catalogue;][]{Hog:00}, and TYC~2076-1387-1 (ck, $V_T = 11.53$).  Both
comparison stars are within 10 arcmin of the variable star (var).  The
comparison star magnitude differences (comp$-$ck) were constant at the
level of 0.013 mag (standard deviation within a night), and 0.010 mag
for the standard deviation of the nightly mean magnitude difference.
The URSA differential magnitude (var$-$comp) of the variable star was
referenced only to the magnitude of the comparison star, comp.  The
resulting 6690 $V$ magnitude differences (var$-$comp) are listed in
Table~\ref{tab:ursa} (without any nightly corrections) and plotted in
Figures~\ref{fig:LCall}--\ref{fig:LCsec} (after the nightly
corrections discussed below have been applied).  The eclipse depths
are 0.68~mag for the primary and 0.66~mag for the secondary. The
precision of the variable star differential magnitudes is about 0.016
mag.

\begin{deluxetable}{ccc}
\tablewidth{0pc}
\tablecaption{URSA differential $V$-band magnitudes of LV~Her.\label{tab:ursa}}
\tablehead{
\colhead{HJD$-2,\!440,\!000$} &
\colhead{$\Delta V$} &
\colhead{Orbital phase}}
\startdata
  51958.87391\dotfill  &  $-$0.195\phs  &    0.15136  \\
  51958.87498\dotfill  &  $-$0.147\phs  &    0.15141  \\
  51958.87603\dotfill  &  $-$0.174\phs  &    0.15147  \\
  51958.87708\dotfill  &  $-$0.179\phs  &    0.15153  \\
  51958.87812\dotfill  &  $-$0.183\phs  &    0.15158  \\ [-1.5ex]
\enddata
\tablecomments{Table~\ref{tab:ursa} is available in its entirety
in the electronic edition of the {\it Astronomical Journal}.  A
portion is shown here for guidance regarding its form and contents.}
\end{deluxetable}

\begin{deluxetable}{ccc}
\tablewidth{0pc}
\tablecaption{NFO differential $V$-band magnitudes of LV~Her.\label{tab:nfo}}
\tablehead{
\colhead{HJD$-2,\!440,\!000$} &
\colhead{$\Delta V$} &
\colhead{Orbital phase}}
\startdata
  53430.93043\dotfill   &    1.003  &    0.99840 \\
  53430.93323\dotfill   &    1.010  &    0.99855 \\
  53430.93602\dotfill   &    1.035  &    0.99870 \\
  53430.93878\dotfill   &    1.055  &    0.99885 \\
  53430.94158\dotfill   &    1.068  &    0.99900 \\ [-1.5ex]
\enddata
\tablecomments{Table~\ref{tab:nfo} is available in its entirety
in the electronic edition of the {\it Astronomical Journal}.  A
portion is shown here for guidance regarding its form and contents.}
\end{deluxetable}

The other telescope we used is the NFO WebScope, a refurbished 24-inch
Group 128 Cassegrain reflector with a 2K $\times$ 2K Kodak CCD camera,
located near Silver City, NM.  Observations consisted of 120 second
exposures also through a Bessell $V$ filter.  LV~Her was observed by
NFO on 123 nights during parts of five observing seasons from 2005
March 1 to 2009 April 24, yielding 2946 observations. Extinction
coefficients were determined nightly from the comparison star
measurements.  They averaged 0.18 mag/airmass.  The same comparison
stars were used as those of the URSA telescope. The comparison star
magnitude differences (comp$-$ck) were constant at the level of 0.007
mag (standard deviation within a night), and 0.020 mag for the
standard deviation of the nightly mean magnitude difference.  The
differential magnitude (var$-$comps) of the variable star was
referenced to the magnitude corresponding to the sum of the
intensities of the comparison star, comp, and the check star, ck.  The
resulting 2946 $V$ magnitude differences (var$-$comps) are listed in
Table~\ref{tab:nfo} (without nightly corrections) and plotted in
Figure~\ref{fig:LCall}--\ref{fig:LCsec} (including nightly
corrections; see below).  The precision of the variable star
differential magnitudes is about 0.010 mag.  We noticed early on
during the observations that the NFO magnitudes showed a small but
significant offset from night to night, on the order of a hundredth of
a magnitude.  The origin of the offset is a variation in responsivity
across the field of view of the NFO combined with imprecise centering
from night to night.  These variations are a well-known effect of the
optics when using wide-field imaging telescopes such as the NFO.  We
have removed most of this variation by using dithered exposures of
open star clusters to measure this variation, fitting a 2-D polynomial
\citep[see][]{Selman:04}, and subtracting the variation during initial
reductions (photometric flat).  The URSA observations, on the other
hand, show this kind of effect to a very much smaller extent. We have
removed these nightly offsets before further analysis by using a
procedure discussed below in \S\thinspace\ref{sec:lcfits}.

Absolute photometry of LV~Her is available in the literature from
several sources, and in several photometric systems including
Str\"omgren, 2MASS, Tycho-2, and Johnson-Cousins. We collect these
measurements in Table~\ref{tab:photometry}. Color indices formed from
these magnitudes can be used to estimate a mean effective temperature
for the binary, which we discuss below in
\S\thinspace\ref{sec:dimensions}.  The interstellar reddening can be
estimated, for example, by comparison with the standard Str\"omgren
indices of \cite{Perry:82}, giving $E(b-y) = 0.007$ mag, or $E(B-V)
\approx 0.01$.  Another estimate is available from the reddening maps
of the sky near the variable star from \cite{Schlegel:98}, which
results in a larger value of $E(B-V) = 0.057$ mag at the estimated
distance of the eclipsing binary. We adopt for further use the
straight average of these two reddening estimates, $E(B-V) = 0.03 \pm
0.03$ mag, with a conservative error.

\begin{deluxetable}{lcc}
\tablewidth{0pc}
\tablecaption{Absolute photometry for LV~Her (combined light).\label{tab:photometry}}
\tablehead{
\colhead{~~~~~Passband~~~~~} &
\colhead{Value} &
\colhead{Reference}
}
\startdata
~~~$y$\dotfill           &  \phantom{\tablenotemark{a}}11.02~$\pm$~0.02\tablenotemark{a}\phn   & 1  \\
~~~$u-b$\dotfill         &   1.552~$\pm$~0.030\tablenotemark{a}  & 1  \\
~~~$v-b$\dotfill         &   0.571~$\pm$~0.020\tablenotemark{a}  & 1  \\
~~~$b-y$\dotfill         &   0.367~$\pm$~0.015\tablenotemark{a}  & 1  \\
~~~$V$\dotfill                            &  11.055~$\pm$~0.040\phn  & 2  \\
~~~$I_{\rm C}$\dotfill                    &  10.367~$\pm$~0.058\phn  & 2  \\
~~~$J$\dotfill         &   9.905~$\pm$~0.018\tablenotemark{a}  & 3  \\
~~~$H$\dotfill         &   9.665~$\pm$~0.025\tablenotemark{a}  & 3  \\
~~~$K_s$\dotfill         &   9.631~$\pm$~0.018\tablenotemark{a}  & 3  \\
~~~$B_T$\dotfill                          &  11.832~$\pm$~0.066\phn  & 4  \\
~~~$V_T$\dotfill                          &  11.045~$\pm$~0.055\phn  & 4  \\ [-1.5ex]
\enddata
\tablecomments{References: 1 - \cite{Hilditch:75}; 2 - TASS \citep{Droege:07};
3 - 2MASS; 4 - Tycho-2 \citep{Hog:00}.}
\tablenotetext{a}{Single measurement obtained out of eclipse.}
\end{deluxetable}

\subsection{Ephemeris}
\label{sec:ephemeris}

\begin{figure}
\epsscale{1.1} 
\plotone{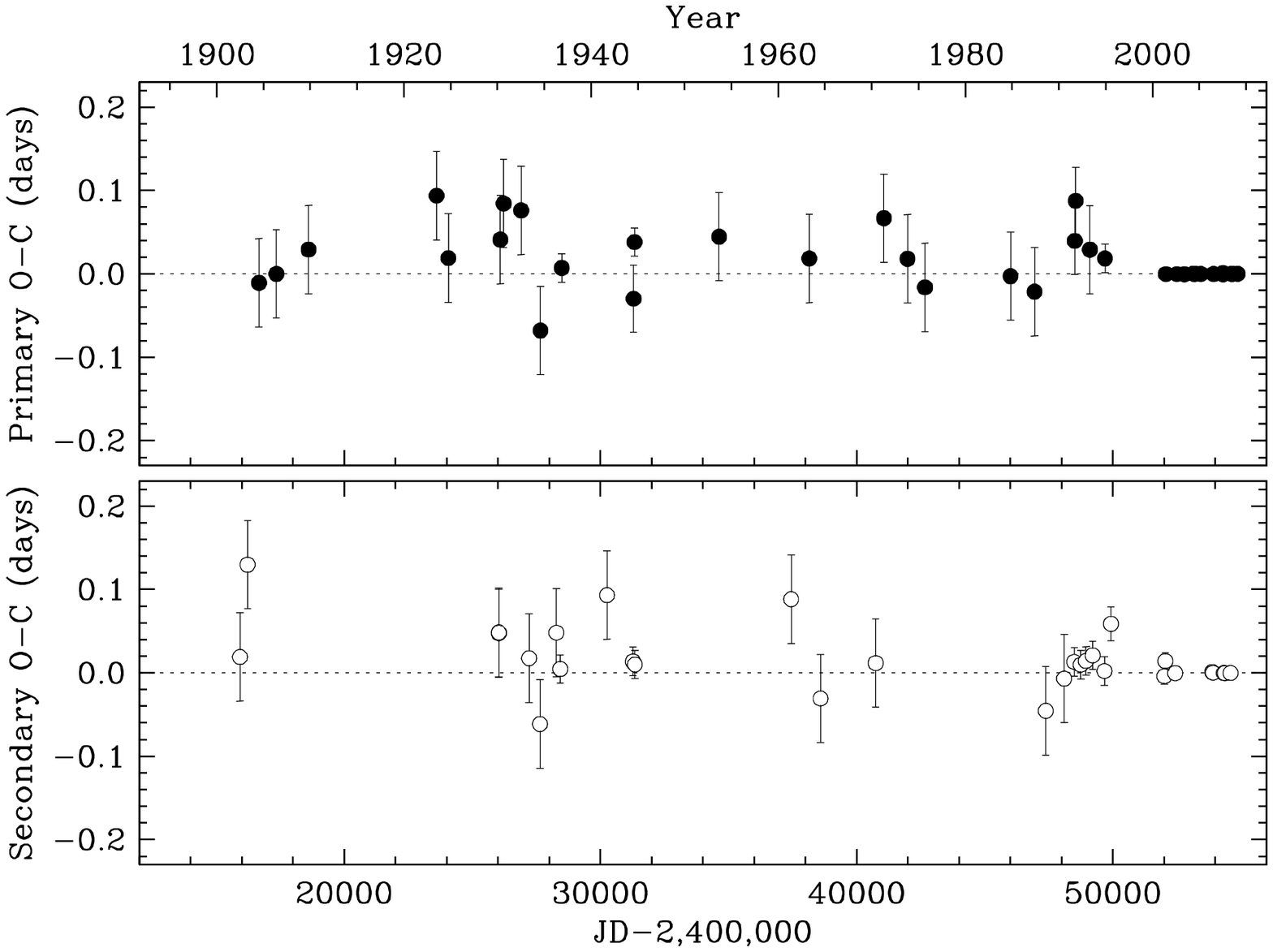}
%
\figcaption[]{Residuals from our fit to the primary and secondary
eclipse timings of LV~Her given in
Table~\ref{tab:minima}.\label{fig:minima}}
\end{figure}

Since the work of \cite{Torres:01}, additional times of minimum for
LV~Her have been reported in the literature, and some recent ones from
our WebScopes are available that have not yet been published. The
previously published ones and these additional times of minimum light
(37 for the primary, 30 for the secondary) are listed in
Table~\ref{tab:minima}. All available minima have been used to
estimate an eclipse ephemeris by a least squares technique. When
available, the uncertainties have been adopted as published. For the
others (visual and photographic measurements), uncertainties have been
estimated by iterations requiring the reduced $\chi^2$ to be near
unity, separately for each class of measurement technique.  For the
visual meaurements we obtain $\sigma = 0.017$~days, and for the
photographic timings $\sigma = 0.053$~days. Separate solutions with
the primary and secondary data give periods that are not significantly
different. For the final ephemeris we enforced a common period, and
obtained:
\begin{eqnarray*}
{\rm Min~I}  &=& 2,\!453,\!652.19147(10) + 18.4359535(19)E \\
{\rm Min~II} &=& 2,\!454,\!165.86045(12) + 18.4359535(19)E
\end{eqnarray*}
where the reference epochs were chosen so as to minimize the
correlations with the period and with each other.  This ephemeris is
the basis of the orbital phases we cite below.  The phase of secondary
eclipse is $0.86235 \pm 0.00001$.  A plot of the $O\!-\!C$ diagram
from this fit is shown in Figure~\ref{fig:minima}, and the residuals
are included in Table~\ref{tab:minima}.  It is quite remarkable how
much the accuracy has improved over the last century.

\begin{deluxetable*}{lcccccc}
\tablewidth{0pc}
\tabletypesize{\scriptsize}
\tablecaption{Times of eclipse for LV~Her.\label{tab:minima}}
\tablehead{
\colhead{HJD$-2,\!400,\!000$} &
\colhead{Year} &
\colhead{Type} &
\colhead{Instrument} &
\colhead{$\sigma$ (d)} &
\colhead{$O\!-\!C$ (d)} &
\colhead{Reference}
}
\startdata
  15929.712\dotfill &   1902.4907   &   2  & pg  &      0.053\pht     &  $+$0.01915  &   1  \\
  16224.798\dotfill &   1903.2986   &   2  & pg  &      0.053\pht     &  $+$0.12989  &   1  \\
  16669.658\dotfill &   1904.5165   &   1  & pg  &      0.053\pht     &  $-$0.01072  &   1  \\
  17351.799\dotfill &   1906.3841   &   1  & pg  &      0.053\pht     &  $-$0.00000  &   1  \\
  18605.473\dotfill &   1909.8165   &   1  & pg  &      0.053\pht     &  $+$0.02916  &   1  \\
  23601.681\dotfill &   1923.4954   &   1  & pg  &      0.053\pht     &  $+$0.09376  &   1  \\
  24062.505\dotfill &   1924.7570   &   1  & pg  &      0.053\pht     &  $+$0.01892  &   1  \\
  26032.643\dotfill &   1930.1510   &   2  & pg  &      0.053\pht     &  $+$0.04762  &   1  \\
  26032.644\dotfill &   1930.1510   &   2  & pg  &      0.053\pht     &  $+$0.04862  &   1  \\
  26090.482\dotfill &   1930.3093   &   1  & pg  &      0.053\pht     &  $+$0.04103  &   1  \\
  26219.577\dotfill &   1930.6628   &   1  & pg  &      0.053\pht     &  $+$0.08436  &   1  \\
  26901.699\dotfill &   1932.5303   &   1  & pg  &      0.053\pht     &  $+$0.07608  &   1  \\
  27212.514\dotfill &   1933.3813   &   2  & pg  &      0.053\pht     &  $+$0.01760  &   1  \\
  27636.462\dotfill &   1934.5420   &   2  & pg  &      0.053\pht     &  $-$0.06134  &   1  \\
  27657.429\dotfill &   1934.5994   &   1  & pg  &      0.053\pht     &  $-$0.06802  &   1  \\
  28281.830\dotfill &   1936.3089   &   2  & pg  &      0.053\pht     &  $+$0.04829  &   1  \\
  28429.2740\dotfill &  1936.7126   &   2  & vis &      0.017\pht     &  $+$0.00466  &   2  \\
  28487.1220\dotfill &  1936.8710   &   1  & vis &      0.017\pht     &  $+$0.00708  &   2  \\
  30254.522\dotfill &   1941.7098   &   2  & pg  &      0.053\pht     &  $+$0.09326  &   1  \\
  31268.42\dotfill &    1944.4857   &   2  & vis &      0.017\pht     &  $+$0.01382  &   3  \\
  31289.35\dotfill &    1944.5431   &   1  & vis &      0.04\phn\pht  &  $-$0.02986  &   3  \\
  31326.29\dotfill &    1944.6442   &   1  & vis &      0.017\pht     &  $+$0.03823  &   3  \\
  31342.16\dotfill &    1944.6876   &   2  & vis &      0.017\pht     &  $+$0.01001  &   3  \\
  34626.332\dotfill &   1953.6792   &   1  & pg  &      0.053\pht     &  $+$0.04456  &   1  \\
  37444.539\dotfill &   1961.3950   &   2  & pg  &      0.053\pht     &  $+$0.08839  &   1  \\
  38147.573\dotfill &   1963.3198   &   1  & pg  &      0.053\pht     &  $+$0.01843  &   1  \\
  38587.449\dotfill &   1964.5242   &   2  & pg  &      0.053\pht     &  $-$0.03072  &   1  \\
  40744.498\dotfill &   1970.4298   &   2  & pg  &      0.053\pht     &  $+$0.01171  &   1  \\
  41060.502\dotfill &   1971.2950   &   1  & pg  &      0.053\pht     &  $+$0.06678  &   1  \\
  41982.251\dotfill &   1973.8186   &   1  & pg  &      0.053\pht     &  $+$0.01810  &   1  \\
  42664.347\dotfill &   1975.6861   &   1  & pg  &      0.053\pht     &  $-$0.01618  &   1  \\
  46001.268\dotfill &   1984.8221   &   1  & pg  &      0.053\pht     &  $-$0.00276  &   1  \\
  46941.483\dotfill &   1987.3963   &   1  & pg  &      0.053\pht     &  $-$0.02139  &   1  \\
  47381.384\dotfill &   1988.6006   &   2  & pg  &      0.053\pht     &  $-$0.04555  &   1  \\
  48100.425\dotfill &   1990.5693   &   2  & pg  &      0.053\pht     &  $-$0.00674  &   1  \\
  48487.60 \dotfill &   1991.6293   &   2  & vis &      0.017\pht     &  $+$0.01328  &   1  \\
  48508.60 \dotfill &   1991.6868   &   1  & vis &      0.04\phn\pht  &  $+$0.03956  &   1  \\
  48545.52 \dotfill &   1991.7879   &   1  & vis &      0.04\phn\pht  &  $+$0.08765  &   1  \\
  48745.70 \dotfill &   1992.3359   &   2  & vis &      0.017\pht     &  $+$0.00989  &   1  \\
  48948.50 \dotfill &   1992.8912   &   2  & vis &      0.017\pht     &  $+$0.01440  &   1  \\
  49098.540\dotfill &   1993.3020   &   1  & pg  &      0.053\pht     &  $+$0.02904  &   1  \\
  49206.61 \dotfill &   1993.5978   &   2  & vis &      0.017\pht     &  $+$0.02105  &   1  \\
  49667.49 \dotfill &   1994.8597   &   2  & vis &      0.017\pht     &  $+$0.00221  &   1  \\
  49688.48 \dotfill &   1994.9171   &   1  & vis &      0.017\pht     &  $+$0.01853  &   1  \\
  49925.65 \dotfill &   1995.5665   &   2  & vis &      0.02\phn\pht  &  $+$0.05886  &   1  \\
  52008.85 \dotfill &   2001.2700   &   2  & ccd &      0.01\phn\pht  &  $-$0.00388  &   1  \\
  52045.74 \dotfill &   2001.3710   &   2  & ccd &      0.01\phn\pht  &  $+$0.01421  &   1  \\
  52066.6993\dotfill &  2001.4283   &   1  & ccd &      0.0008\phn    &  $-$0.00017  &   1  \\
  52432.88040\dotfill & 2002.4309   &   2  & ccd &      0.00050       &  $-$0.00042  &   \phn4* \\
  52490.72613\dotfill & 2002.5893   &   1  & ccd &      0.00020       &  $-$0.00027  &   4  \\
  52785.7012\dotfill &  2003.3969   &   1  & ccd &      0.0010\phn    &  $-$0.00046  &   5  \\
  53154.4210\dotfill &  2004.4064   &   1  & ccd &      0.0017\phn    &  $+$0.00027  &   6  \\
  53154.4212\dotfill &  2004.4064   &   1  & ccd &      0.0015\phn    &  $+$0.00047  &   6  \\
  53209.7288\dotfill &  2004.5578   &   1  & ccd &      0.0004\phn    &  $+$0.00021  &   7  \\
  53430.95992\dotfill & 2005.1635   &   1  & ccd &      0.00016       &  $-$0.00011  &   8  \\
  53870.8862\dotfill &  2006.3679   &   2  & ccd &      0.0005\phn    &  $+$0.00101  &   \phn9* \\
  53907.7573\dotfill &  2006.4689   &   2  & ccd &      0.0002\phn    &  $+$0.00020  &   9  \\
  53928.7308\dotfill &  2006.5263   &   1  & ccd &      0.0003\phn    &  $+$0.00002  &   9  \\
  54297.44944\dotfill & 2007.5358   &   1  & ccd &      0.00040       &  $-$0.00041  &   10 \\
  54297.4498\dotfill &  2007.5358   &   1  & ccd &      0.0008\phn    &  $-$0.00005  &   11 \\
  54297.4509\dotfill &  2007.5358   &   1  & ccd &      0.0004\phn    &  $+$0.00105  &   12 \\
  54331.7839\dotfill &  2007.6298   &   2  & ccd &      0.0003\phn    &  $-$0.00013  &   13 \\
  54368.6555\dotfill &  2007.7307   &   2  & ccd &      0.0003\phn    &  $-$0.00043  &   13 \\
  54589.8873\dotfill &  2008.3364   &   2  & ccd &      0.0002\phn    &  $-$0.00008  &   13 \\
  54647.7328\dotfill &  2008.4948   &   1  & ccd &      0.0006\phn    &  $-$0.00016  &   13 \\
  54647.7332\dotfill &  2008.4948   &   1  & ccd &      0.0002\phn    &  $+$0.00024  &   13 \\
  54868.9645\dotfill &  2009.1005   &   1  & ccd &      0.0007\phn    &  $+$0.00010  &   13 \\ [-1.5ex]
\enddata
\tablecomments{Type: 1 = primary, 2 = secondary.  Instrument: pg =
photographic, vis = visual, ccd = CCD. References:
1 - \cite{Torres:01};
2 - Variable Star and Exoplanet Section of the Czech Astronomical Society
{\tt{(http://var2.astro.cz/EN/brno/eclipsing\_binaries.php)}};
3 - \cite{Zessewitsch:54};
4 - \cite{Lacy:02};
5 - \cite{Lacy:03};
6 - \cite{Hubscher:05};
7 - \cite{Lacy:04};
8 - \cite{Lacy:06};
9 - \cite{Lacy:07};
10 - \cite{Brat:07};
11 - \cite{Hubscher:08};
12 - \cite{Diethelm:08};
13 - This paper;
* - Remeasurement.
}
\end{deluxetable*}

\subsection{Spectroscopy}
\label{sec:spectroscopy}

LV~Her was placed on the observing list at the Harvard-Smithsonian
Center for Astrophysics (CfA) on 1991 May 31st, and was monitored
spectroscopically for exactly 11 years with an echelle instrument on
the 1.5m Tillinghast reflector at the F.\ L.\ Whipple Observatory on
Mount Hopkins, Arizona. We collected 42 spectra with a photon-counting
intensified Reticon detector \citep{Latham:85, Latham:92} in a single
order 45~\AA\ wide centered near 5187~\AA. The main features in this
spectral window are the lines of the \ion{Mg}{1}~b triplet. The
resolving power provided by this setup is
$\lambda/\Delta\lambda\approx 35,\!000$. Two additional observations,
for a total of 44, were obtained with a nearly identical system on the
1.5m Wyeth reflector at the Oak Ridge Observatory (Harvard,
Massachusetts). The signal-to-noise ratios of these spectra range from
13 to 28 per resolution element of 8.5~\kms.

Radial velocities for the two components were derived using TODCOR, a
two-dimensional cross-correlation technique \citep{Zucker:94}. This
method uses two templates, one for each component of the binary, which
we selected from a large library of synthetic spectra based on model
atmospheres by R.\ L.\ Kurucz \citep[see][]{Latham:02}. These
templates have been calculated for a wide range of effective
temperatures ($T_{\rm eff}$), surface gravities ($\log g$), rotational
velocities ($v \sin i$ when seen in projection), and metallicities
([m/H]). Following \cite{Torres:02} the optimum templates for each
star were determined by means of extensive grids of cross-correlations
with TODCOR, seeking to maximize the average correlation weighted by
the strength of each exposure. Because of the strong correlation
between temperature, surface gravity, and metallicity in our narrow
spectral window, we initially assumed solar metallicity, and surface
gravities of $\log g = 4.0$ for both components, close to the values
determined from a preliminary analysis.  The projected rotational
velocities we obtained are $v \sin i = 13 \pm 1$~\kms\ for both stars,
and the best-fit temperatures were near solar, with a difference of
only 20--30~K between the primary and secondary. For the radial
velocity measurements we adopted $T_{\rm eff} = 5750$~K.  In
\S\thinspace\ref{sec:dimensions} we describe experiments in which we
extended the grids of correlations to other compositions in order to
fine-tune the temperatures and attempt to constrain the
metallicity. These minor changes in the stellar parameters have no
effect on the radial velocities.  Typical errors for our measurements
are 0.8~\kms\ for the primary and 1.1~\kms\ for the secondary.  The
stability of the zero-point of the CfA velocity system was monitored
by means of exposures of the dusk and dawn sky, and small systematic
run-to-run corrections were applied in the manner described by
\cite{Latham:92}.  We determined also the light ratio directly from
our spectra following \cite{Zucker:94}, and obtained $\ell_2/\ell_1 =
0.95 \pm 0.03$ at the mean wavelength of our observations, which is
essentially also the ratio in the visual band given that the stars are
nearly identical in temperature.

\begin{deluxetable}{lc}[b]
\tablewidth{0pc}
\tablecaption{Spectroscopic orbital solution for LV~Her.\label{tab:specorbit}}
\tablehead{
\colhead{
\hfil~~~~~~~~~~~~~Parameter~~~~~~~~~~~~~~} & \colhead{Value}}
\startdata
\multicolumn{2}{l}{Adjusted quantities\hfil} \\
~~~~$P$ (days)\tablenotemark{a}\dotfill        &  18.4359535                   \\
~~~~$T_{\rm I}$ (HJD$-$2,400,000)\tablenotemark{a}\dotfill     & 53,652.19147  \\
~~~~$K_1$ (\kms)\dotfill                       &   67.24~$\pm$~0.19\phn        \\
~~~~$K_2$ (\kms)\dotfill                       &   68.59~$\pm$~0.27\phn        \\
~~~~$\gamma$ (\kms)\dotfill                    & $-10.278$~$\pm$~0.094\phs\phn \\
~~~~$e$\dotfill                                & 0.61273~$\pm$~0.00073         \\
~~~~$\omega$ (deg)\dotfill                     & 352.20~$\pm$~0.24\phn\phn     \\
\multicolumn{2}{l}{Derived quantities\hfil} \\
~~~~$M_1\sin^3 i$ ($M_{\sun}$)\dotfill         &  1.193~$\pm$~0.010            \\
~~~~$M_2\sin^3 i$ ($M_{\sun}$)\dotfill         &  1.1697~$\pm$~0.0080          \\
~~~~$q\equiv M_2/M_1$\dotfill                  &  0.9803~$\pm$~0.0047          \\
~~~~$a_1\sin i$ (10$^6$ km)\dotfill            &  13.471~$\pm$~0.037\phn       \\
~~~~$a_2\sin i$ (10$^6$ km)\dotfill            &  13.743~$\pm$~0.054\phn       \\
~~~~$a \sin i$ ($R_{\sun}$)\dotfill            &  39.119~$\pm$~0.095\phn       \\
\multicolumn{2}{l}{Other quantities pertaining to the fit\hfil} \\
~~~~$N_{\rm obs}$\dotfill                      & 44                            \\
~~~~Time span (days)\dotfill                   &  4018.0                       \\
~~~~$\sigma_1$ (\kms)\dotfill                  & 0.76                          \\
~~~~$\sigma_2$ (\kms)\dotfill                  & 1.10                          \\ [-1.5ex]
\enddata
\tablenotetext{a}{Ephemeris adopted from \S\,\ref{sec:ephemeris}.}
\end{deluxetable}

\begin{figure}[b]
\epsscale{1.1} 
\plotone{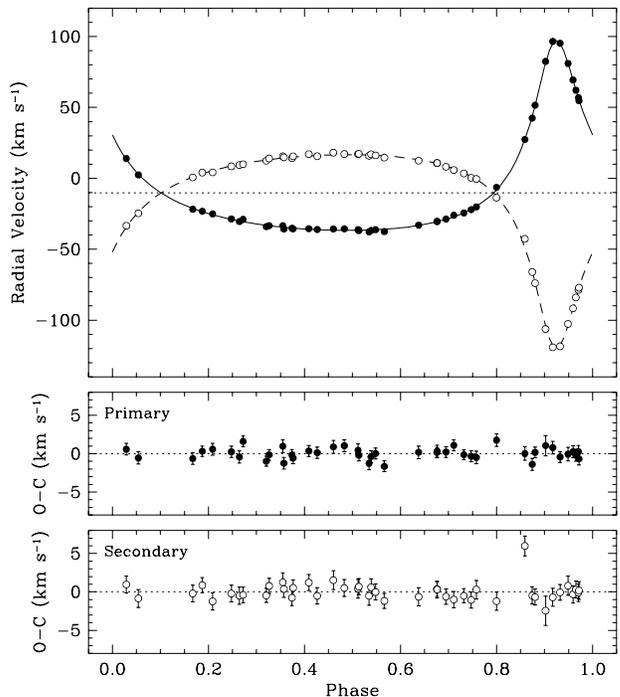}
%
\figcaption[]{Radial-velocity measurements for LV~Her (filled circles
for the primary, open for the secondary) along with the curves
computed from our orbital solution. Phase 0.0 corresponds to the time
of primary eclipse. The dotted line represents the velocity of the
center of mass, and the error bars are smaller than the symbol
size. $O\!-\!C$ residuals for the primary and secondary are shown in
the bottom panels.\label{fig:sb2orbit}}
\end{figure}

\begin{deluxetable*}{lccccccc}[b]
\tablewidth{0pc}
\tablecaption{Radial velocity measurements of LV~Her.\label{tab:rvs}}
\tablehead{\colhead{HJD} & 
\colhead{$RV_1$} & 
\colhead{$RV_2$} & 
\colhead{$\sigma_1$} & 
\colhead{$\sigma_2$} & 
\colhead{$(O\!-\!C)_1$} & 
\colhead{$(O\!-\!C)_2$} & 
\colhead{} \\
\colhead{\hbox{~~(2,400,000$+$)~~}} & 
\colhead{(\kms)} & 
\colhead{(\kms)} & 
\colhead{(\kms)} & 
\colhead{(\kms)} & 
\colhead{(\kms)} & 
\colhead{(\kms)} & 
\colhead{Phase}}
\startdata
 48407.8794\dotfill &  $-$36.77  & $+$16.94  &  0.75  &  1.09  &  $-$0.38 & $+$0.59  &   0.5389 \\
 48428.8481\dotfill &  $-$30.28  & $+$10.75  &  0.77  &  1.12  &  $+$0.32 & $+$0.30  &   0.6763 \\
 48435.8126\dotfill &   \phn$+$2.41  & $-$24.64  &  0.81  &  1.17  &  $-$0.56 & $-$0.85  &   0.0541 \\
 48459.7950\dotfill &  $-$33.43  & $+$15.58  &  0.84  &  1.22  &  $+$0.97 & $+$1.25  &   0.3549 \\
 48461.7375\dotfill &  $-$35.70  & $+$18.08  &  0.85  &  1.23  &  $+$0.88 & $+$1.53  &   0.4603 \\
 48462.6830\dotfill &  $-$36.20  & $+$17.06  &  0.84  &  1.22  &  $+$0.44 & $+$0.44  &   0.5116 \\
 51709.7370\dotfill &  $-$33.05  & $+$12.51  &  0.82  &  1.19  &  $+$0.18 & $-$0.62  &   0.6377 \\
 51710.7867\dotfill &  $-$28.71  &  \phn$+$8.12  &  0.69  &  1.00  &  $+$0.21 & $-$0.61  &   0.6947 \\
 51711.7546\dotfill &  $-$22.10  &  \phn$+$0.39  &  0.72  &  1.04  &  $-$0.34 & $-$1.05  &   0.7472 \\
 51712.7284\dotfill &   \phn$-$6.38  & $-$13.66  &  0.81  &  1.18  &  $+$1.76 & $-$1.20  &   0.8000 \\
 51715.7770\dotfill &  $+$62.12  & $-$84.03  &  0.77  &  1.11  &  $-$0.20 & $+$0.31  &   0.9653 \\
 51740.7610\dotfill &  $-$34.04  & $+$12.50  &  0.64  &  0.93  &  $-$0.98 & $-$0.46  &   0.3205 \\
 51741.7446\dotfill &  $-$35.18  & $+$14.19  &  0.72  &  1.04  &  $-$0.18 & $-$0.75  &   0.3739 \\
 51742.7170\dotfill &  $-$36.06  & $+$15.62  &  0.74  &  1.07  &  $+$0.12 & $-$0.52  &   0.4266 \\
 51744.7057\dotfill &  $-$37.70  & $+$15.90  &  0.83  &  1.21  &  $-$1.26 & $-$0.51  &   0.5345 \\
 51799.6219\dotfill &  $-$36.83  & $+$17.30  &  0.73  &  1.05  &  $-$0.20 & $+$0.69  &   0.5132 \\
 51800.6010\dotfill &  $-$37.55  & $+$14.68  &  0.69  &  1.00  &  $-$1.65 & $-$1.17  &   0.5663 \\
 51802.6229\dotfill &  $-$30.48  & $+$10.83  &  0.66  &  0.95  &  $+$0.14 & $+$0.36  &   0.6760 \\
 51803.6558\dotfill &  $-$24.42  &  \phn$+$3.48  &  0.64  &  0.92  &  $-$0.15 & $-$0.51  &   0.7320 \\
 51831.5984\dotfill &  $-$28.65  &  \phn$+$8.50  &  0.75  &  1.08  &  $+$0.24 & $-$0.21  &   0.2477 \\
 51833.6157\dotfill &  $-$35.73  & $+$14.79  &  0.75  &  1.09  &  $-$1.25 & $+$0.38  &   0.3571 \\
 51834.5731\dotfill &  $-$35.52  & $+$17.04  &  0.72  &  1.04  &  $+$0.34 & $+$1.22  &   0.4091 \\
 51858.5764\dotfill &  $-$26.02  &  \phn$+$5.88  &  0.73  &  1.06  &  $+$1.09 & $-$1.01  &   0.7110 \\
 51861.5870\dotfill &  $+$42.50  & $-$66.00  &  0.77  &  1.12  &  $-$1.39 & $-$0.46  &   0.8743 \\
 51973.9772\dotfill &  $+$56.88  & $-$78.61  &  0.79  &  1.14  &  $+$0.26 & $-$0.09  &   0.9706 \\
 51977.9683\dotfill &  $-$23.19  &  \phn$+$4.08  &  0.68  &  0.98  &  $+$0.30 & $+$0.88  &   0.1871 \\
 52006.9280\dotfill &  $-$20.18  &  \phn$-$0.38  &  0.83  &  1.21  &  $-$0.49 & $+$0.30  &   0.7579 \\
 52009.8622\dotfill &  $+$96.55  &$-$119.15\phn  &  0.81  &  1.18  &  $+$0.79 & $-$0.70  &   0.9171 \\
 52010.8700\dotfill &  $+$54.74  & $-$77.11  &  0.80  &  1.16  &  $-$0.68 & $+$0.19  &   0.9717 \\
 52011.9261\dotfill &  $+$13.95  & $-$33.43  &  0.76  &  1.10  &  $+$0.58 & $+$0.97  &   0.0290 \\
 52032.9120\dotfill &  $-$21.81  &  \phn$+$0.64  &  0.75  &  1.09  &  $-$0.63 & $-$0.20  &   0.1673 \\
 52034.8450\dotfill &  $-$28.93  &  \phn$+$9.98  &  0.71  &  1.04  &  $+$1.59 & $-$0.39  &   0.2722 \\
 52035.8449\dotfill &  $-$33.46  & $+$14.00  &  0.70  &  1.02  &  $-$0.15 & $+$0.78  &   0.3264 \\
 52039.9325\dotfill &  $-$36.23  & $+$16.20  &  0.73  &  1.05  &  $+$0.02 & $-$0.01  &   0.5481 \\
 52102.6290\dotfill &  $+$80.99  &$-$102.64\phn  &  0.87  &  1.26  &  $-$0.07 & $+$0.81  &   0.9489 \\
 52138.6401\dotfill &  $+$82.46  &$-$106.28\phn  &  1.31  &  1.91  &  $+$1.04 & $-$2.45  &   0.9022 \\
 52156.6725\dotfill &  $+$51.60  & $-$73.90  &  0.71  &  1.04  &  $+$0.17 & $-$0.67  &   0.8803 \\
 52157.6303\dotfill &  $+$95.25  &$-$118.44\phn  &  0.70  &  1.01  &  $-$0.44 & $-$0.06  &   0.9323 \\
 52211.5883\dotfill &  $+$27.39  & $-$42.70  &  0.89  &  1.30  &  $+$0.01 & $+$5.99  &   0.8590 \\
 52360.9241\dotfill &  $+$69.40  & $-$91.63  &  0.78  &  1.13  &  $+$0.24 & $-$0.31  &   0.9593 \\
 52420.8295\dotfill &  $-$25.10  &  \phn$+$4.20  &  0.77  &  1.11  &  $+$0.58 & $-$1.23  &   0.2087 \\
 52421.8579\dotfill &  $-$30.47  &  \phn$+$9.39  &  0.78  &  1.13  &  $-$0.43 & $-$0.49  &   0.2645 \\
 52423.9196\dotfill &  $-$35.64  & $+$15.54  &  0.73  &  1.06  &  $-$0.57 & $+$0.53  &   0.3763 \\
 52425.8862\dotfill &  $-$35.66  & $+$17.17  &  0.77  &  1.11  &  $+$1.03 & $+$0.51  &   0.4830 \\ [-1.5ex]
\enddata
\tablecomments{Radial velocities are in the heliocentric frame, and
include all corrections described in the text.}
\end{deluxetable*}

Although TODCOR significantly reduces systematic errors in the radial
velocities caused by line blending, residual effects can remain as a
result of shifts of the spectral lines in and out of our narrow
spectral window as a function of orbital phase.  We investigated these
effects by means of numerical simulations similar to those described
by \cite{Latham:96} \citep[see also][]{Torres:97, Torresetal:00a}. We
generated synthetic composite spectra matching our observations by
combining copies of the primary and secondary templates used above,
shifted to the appropriate velocities for each of the exposures as
predicted by a preliminary orbital solution, and scaled to the
observed light ratio. These synthetic observations were then processed
with TODCOR in exactly the same way as the real spectra, and the
resulting velocities were compared with the input shifts. The
differences for LV~Her were typically well under 0.5~\kms\ for both
stars, but were nevertheless applied as corrections to the raw
velocities. They affect the masses at the level of 1.3\% for the
primary and 0.8\% for the secondary, which are similar to the formal
errors. The corrected measurements are listed in Table~\ref{tab:rvs}
along with their uncertainties. Similar adjustments based on the same
simulations were made to the light ratio, and are already included in
the value reported above.

Our orbital solution is shown in Figure~\ref{fig:sb2orbit}, and was
derived holding the ephemeris fixed according to
\S\thinspace\ref{sec:ephemeris}. The residuals are shown as well, and
are listed in Table~\ref{tab:rvs}. The orbital elements and derived
quantities (minimum masses, semimajor axes, etc.) are given in
Table~\ref{tab:specorbit}.

\section{Modeling of the photometric observations}
\label{sec:lcfits}

\begin{deluxetable*}{lccc}[b]
\tablewidth{0pc}
\tablecaption{Photometric orbital solutions for LV~Her.\label{tab:lcfit}}
\tablehead{
\colhead{~~~~~~~~~~Parameter~~~~~~~~~~} &
\colhead{URSA} &
\colhead{NFO}  &
\colhead{Adopted}
}
\startdata
~~~~$J_2$\dotfill                  & 0.967~$\pm$~0.004          & 0.983~$\pm$~0.003          & 0.9772~$\pm$~0.0080        \\
~~~~$r_1+r_2$\dotfill            & 0.06880~$\pm$~0.00009      & 0.06806~$\pm$~0.00006      & 0.06829~$\pm$~0.00037      \\
~~~~$k \equiv r_2/r_1$\dotfill   & 0.986~$\pm$~0.015          & 0.961~$\pm$~0.009          & 0.968~$\pm$~0.012          \\
~~~~$r_1$\dotfill                & 0.03465~$\pm$~0.00026      & 0.03471~$\pm$~0.00015      & 0.03470~$\pm$~0.00028      \\
~~~~$r_2$\dotfill                & 0.03415~$\pm$~0.00026      & 0.03335~$\pm$~0.00016      & 0.03357~$\pm$~0.00028      \\
~~~~$i$ (deg)\dotfill            & 89.489~$\pm$~0.010\phn     & 89.509~$\pm$~0.009\phn     & 89.500~$\pm$~0.010\phn     \\
~~~~$e \cos\omega$\dotfill       & 0.60662~$\pm$~0.00011      & 0.60739~$\pm$~0.00006      & 0.60721~$\pm$~0.00038      \\
~~~~$e \sin\omega$\dotfill       & $-$0.0938~$\pm$~0.0019\phs & $-$0.0796~$\pm$~0.0012\phs & $-$0.0836~$\pm$~0.0071\phs \\
~~~~$e$\dotfill                  & 0.61383~$\pm$~0.00018      & 0.61259~$\pm$~0.00010      & 0.61288~$\pm$~0.00062      \\
~~~~$\omega$ (deg)\dotfill       & 351.21~$\pm$~0.18\phn\phn  & 352.53~$\pm$~0.12\phn\phn  & 352.12~$\pm$~0.66\phn\phn  \\
~~~~$\ell_1$\dotfill             & 0.516~$\pm$~0.007          & 0.524~$\pm$~0.004          & 0.5220~$\pm$~0.0040        \\
~~~~$\ell_2$\dotfill             & 0.484~$\pm$~0.007          & 0.476~$\pm$~0.004          & 0.4780~$\pm$~0.0040        \\
~~~~$(\ell_2/\ell_1)_V$\dotfill  & 0.938~$\pm$~0.024          & 0.908~$\pm$~0.014          & 0.916~$\pm$~0.015          \\
~~~~$u_1 = u_2$\dotfill          & 0.53~$\pm$~0.04            & 0.55~$\pm$~0.04            & 0.54~$\pm$~0.03            \\ 
~~~~$N_{\rm obs}$\dotfill        & 6690                       & 2946                       & \nodata                    \\
~~~~Time span (days)\dotfill     & 2970.9                     & 1514.9                     & \nodata                    \\
~~~~$\sigma_V$ (mmag)\dotfill    & 12.49                      & 6.42                       & \nodata                    \\[-1.5ex]
\enddata
\end{deluxetable*}

In order to remove the small nightly offsets in the differential
photometry, we performed preliminary fits of the URSA and NFO light
curves by using the Nelson-Davis-Etzel model \citep{Popper:81,
Etzel:81} as implemented in the EBOP code. This model is well suited
for well-detached systems such as LV~Her.  The nightly residuals,
which were typically less than 0.02 mag and were uncorrelated with
phase, were removed from each light curve. This improved the residuals
of the URSA data by about 27\%, and those from NFO by a more
significant 42\%.  The corrected light curves were then fitted by
using the JKTEBOP code of \cite{Southworth:07}, based on the same
model, which allows for more realistic estimates of the uncertainties.
The adjustable parameters are the central surface brightness $J_2$ of
the secondary relative to the primary, the inclination angle $i$, the
sum of the relative radii $r_1+r_2$, the ratio of the radii $k \equiv
r_2/r_1$, the eccentricity factors $e \cos\omega$ and $e \sin \omega$,
a phase shift, and the magnitude at quadrature. A linear
limb-darkening law was adopted, consistent with our experience that
with the amount and precision of our data, nonlinear laws do not
improve the accuracy of the fits significantly
\citep[e.g.,][]{Lacy:05, Lacy:08}. The coefficient $u$ was left free,
and constrained to be the same for the two stars. The gravity
brightening exponent was set to 0.35 for both stars, based on the
calculations by \cite{Claret:98} and the mean temperature and surface
gravity of the components. The mass ratio $q$ was adopted from the
spectroscopy.

Separate solutions for the URSA and NFO photometry are given in
Table~\ref{tab:lcfit}. The uncertainties for the fitted parameters
were evaluated by Monte Carlo simulation with JKTEBOP.  Tests
indicated third light was not significant, and it was therefore set to
zero.  The last column of the table gives the adopted fit (weighted
average), with uncertainties that account for the difference between
URSA and NFO. The uncertainties for the individual radii were derived
following the prescription recommended by \cite{Torresetal:00b}, based
on the errors for $r_1+r_2$ and $k$.  The synthetic curve
corresponding to the adopted solution is displayed in
Figures~\ref{fig:LCall}--\ref{fig:LCsec} along with the observations.
The eccentricity and $\omega$ from the light curves are in excellent
agreement with the spectroscopic values in Table~\ref{tab:specorbit},
and the photometric light ratio $(\ell_2/\ell_1)_V = 0.916 \pm 0.015$
is consistent with the estimate of $\ell_2/\ell_1 = 0.95 \pm 0.03$
from the CfA spectra.  The well-detached stars are essentially
spherical, and both eclipses are partial, with 88.4\% of the primary
light blocked at the primary minimum, and 95.6\% of the secondary
light covered at the secondary minimum.

We note that the linear limb-darkening coefficient we determined, $u =
0.54 \pm 0.03$, is consistent with theoretical value of 0.568 for the
visual band according to \cite{vanHamme:93}, but is significantly
smaller than the value 0.658 from the tables by \cite{Claret:00}, for
a star of this temperature and surface gravity. Similar discrepancies
have been found in other systems \citep[see, e.g.,][]{Claret:08}.  As
a test we repeated the light curve fits holding the limb-darkening
coefficient fixed at this last theoretical value. The change in
$r_1+r_2$ was only +0.15\%, the individual radii increased by less
than 0.1\%, and the inclination angle decreased by $0\fdg065$. These
changes do not impact the absolute masses or radii at a significant
level, compared to other sources of uncertainty.

\section{Absolute dimensions}
\label{sec:dimensions}

The photometric orbit and the spectroscopic orbit may be combined to
yield the absolute masses and radii of the binary stars' components.
These are given in Table~\ref{tab:dimensions}, along with other
properties described below. The masses and radii are determined to
0.9\% or better for both stars. The tests and checks for systematics
described above suggest that they are also accurate at this level.

The ratio of the effective temperatures is very well constrained from
the light curves through the central surface brightness parameter
$J_2$, according to which the stars differ by only about 30~K
\citep[][Table~1]{Popper:80}. A similar difference was obtained from
the CfA spectra.  The \emph{absolute} temperatures of the stars may in
principle be estimated also from the spectroscopy, though in practice
this is difficult with the material at hand because of the correlation
with metallicity mentioned earlier.  We first determined the magnitude
of this dependence by repeating the grids of correlations described in
\S\thinspace\ref{sec:spectroscopy} for metallicities between [m/H] $=
-1.0$ and [m/H] $= +0.5$ in steps of 0.5 dex, and surface gravities of
$\log g = 4.0$ and 4.5. At each composition we determined the primary
and secondary temperatures interpolated to the surface gravities
indicated in Table~\ref{tab:dimensions}, as well as the
luminosity-weighted mean temperature for the system (where the weights
depend only on the radius ratio and temperature ratio that come from
the light curve fits). An additional constraint on the mean system
temperature is available from the absolute photometry of LV~Her
presented earlier, through color indices and color-temperature
calibrations (including metallicity terms). Equating the two mean
temperatures then allows to solve for the metallicity.  With the
photometry listed in Table~\ref{tab:photometry} we formed 7 different
color indices, we de-reddened them as indicated in
\S\thinspace\ref{sec:photometry}, and we made use of three different
calibrations by \cite{Ramirez:05}, \cite{Casagrande:06}, and
\cite{Gonzalez:09}. While the 7 indices are not independent, they are
useful in providing an idea of the internal consistency of the
results. The agreement between the different indices and different
calibrations is excellent, generally within 100~K.  The metallicity
dependence of the photometric temperatures is small, but was
nevertheless accounted for by iterations with the spectroscopic
results. We obtained a mean system temperature of $T_{\rm eff} = 6050
\pm 140$~K and [m/H] $= +0.08 \pm 0.21$, where the errors are
dominated by the uncertain reddening. The primary and secondary
temperatures are then $6060 \pm 150$~K and $6030 \pm 150$~K,
respectively, though of course $\Delta T_{\rm eff}$ is known much more
precisely. These temperatures correspond to a spectral type of
approximately F9 \citep[e.g.,][]{Popper:80}.

LV~Her does not have an entry in the {\it Hipparcos\/} catalog
\citep{Perryman:97}.  The distance to the system is estimated here as
$352 \pm 24$~pc, based on the bolometric luminosities, the apparent
system magnitude of $V = 11.02 \pm 0.02$ (Table~\ref{tab:photometry}),
the adopted extinction of $3.1 \times E(B-V)$, and bolometric
corrections from \cite{Flower:96}. Separate distance estimates for
each component agree nearly perfectly, showing the internal
consistency of the results.

Also listed in Table~\ref{tab:dimensions} are the rotational
velocities (projected along the line of sight) expected if the stars
were pseudo-synchronized \citep[see][]{Hut:81}, as well as those that
correspond to synchronization with the angular velocity at periastron,
for comparison with the measured $v \sin i$ values. We discuss these
below.

\begin{deluxetable}{lcc}[b]
\tablewidth{0pc}
\tablecaption{Physical properties of LV~Her.\label{tab:dimensions}}
\tablehead{
\colhead{~~~~~~~~~~Parameter~~~~~~~~~~} &
\colhead{Primary} &
\colhead{Secondary}}
\startdata
Mass ($M_{\sun}$)\dotfill            &   1.193~$\pm$~0.010   &     1.1698~$\pm$~0.0081 \\
Radius ($R_{\sun}$)\dotfill          &   1.358~$\pm$~0.012   &     1.313~$\pm$~0.011 \\
$\log g$ (cgs)\dotfill               &   4.2493~$\pm$~0.0082 &     4.2695~$\pm$~0.0081 \\
Temperature (K)\dotfill              &   6060~$\pm$~150\phn  &     6030~$\pm$~150\phn \\
$\log L$ ($L_{\sun}$)\dotfill        &   0.349~$\pm$~0.044   &     0.311~$\pm$~0.044 \\
$M_{\rm bol}$ (mag)\dotfill          &   3.86~$\pm$~0.11     &     3.95~$\pm$~0.11 \\
$BC_V$\tablenotemark{a}\dotfill                       & $-$0.04~$\pm$~0.10\phs  & $-$0.04~$\pm$~0.10\phs \\
$M_V$ (mag)\tablenotemark{b}\dotfill                  &   3.90~$\pm$~0.16     &     4.00~$\pm$~0.16 \\
$a$ ($R_{\sun}$)\dotfill             &  \multicolumn{2}{c}{39.120~$\pm$~0.095\phn} \\
Distance (pc)\dotfill                &  \multicolumn{2}{c}{352~$\pm$~24\phn} \\
$[{\rm m/H}]$\dotfill                &  \multicolumn{2}{c}{+0.08~$\pm$~0.21\phs} \\
$v \sin i$ (\kms)\tablenotemark{c}\dotfill            &   13~$\pm$~1\phn      &     13~$\pm$~1\phn   \\
$v_{\rm psync} \sin i$ (\kms)\tablenotemark{d}\dotfill &   16.0~$\pm$~0.1\phn  &     15.5~$\pm$~0.1\phn \\
$v_{\rm peri} \sin i$ (\kms)\tablenotemark{e}\dotfill &   19.6~$\pm$~0.2\phn  &     19.0~$\pm$~0.2\phn \\[-1.5ex]
\enddata
\tablenotetext{a}{$BC_V$ is taken from \cite{Flower:96}, and the uncertainty includes the contribution
from the temperature as well as an additional 0.10 mag added in quadrature.}
\tablenotetext{b}{The bolometric magnitude adopted for the Sun is $M_{\rm bol}^{\sun} = 4.732$, for
consistency with the zero-point of the bolometric corrections from \cite{Flower:96}.}
\tablenotetext{c}{Value measured spectroscopically.}
\tablenotetext{d}{Projected pseudo-synchronous rotational velocity.}
\tablenotetext{e}{Projected synchronous rotational velocity at periastron.}
\end{deluxetable}

\section{Comparison with theory}

The high accuracy and precision of our measurements for LV~Her offers
the opportunity to test various aspects of theoretical modeling. In
this particular case, knowledge of the chemical composition, even
though the precision is not as high, is especially interesting since
it is rarely available for eclipsing binaries and it eliminates one of
the free parameters in the comparison with theory.

\subsection{Stellar evolution theory}
\label{sec:evolution}

\begin{figure}[b]
\epsscale{1.1} 
\plotone{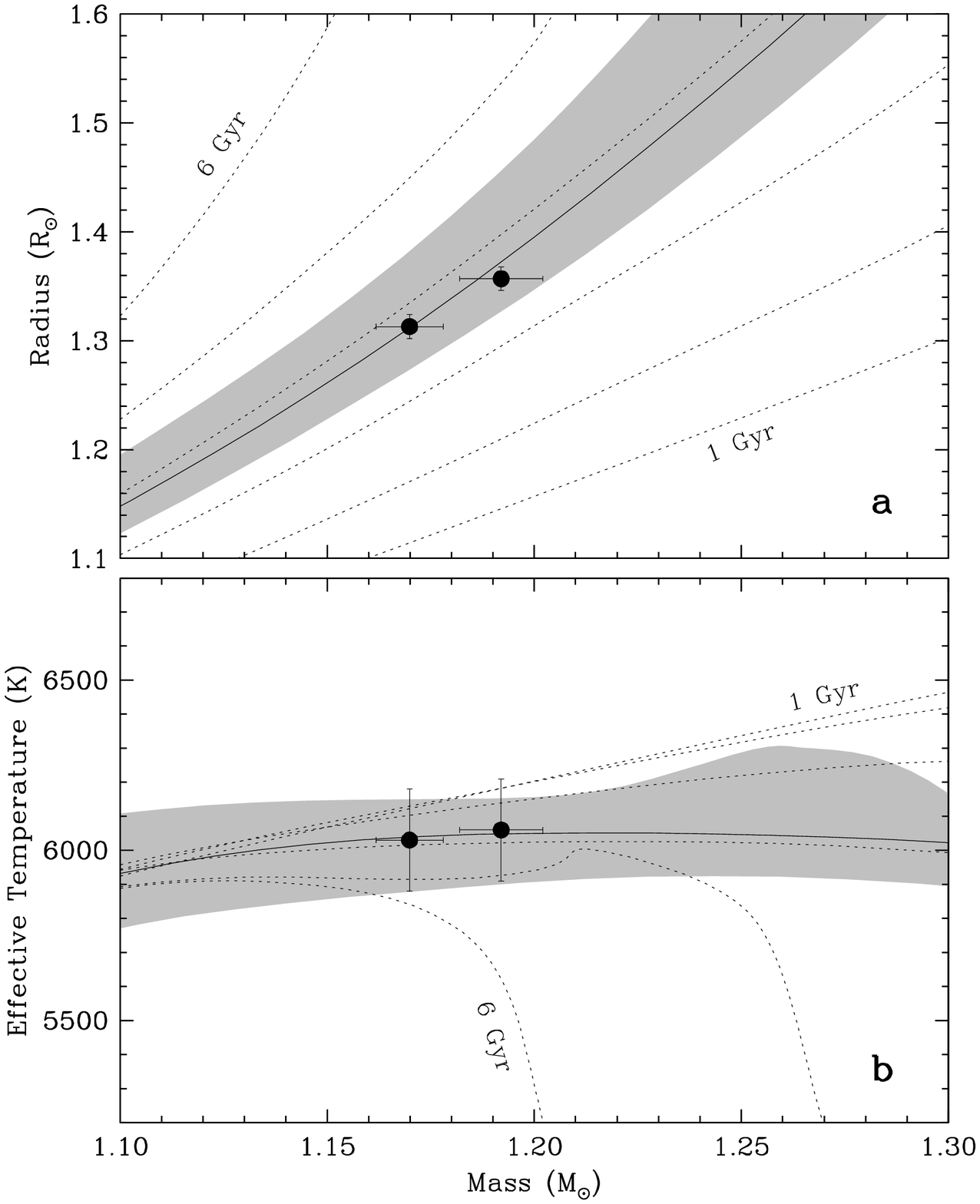}     
%
\figcaption[]{Measurements for LV~Her compared against the Yonsei-Yale
models \citep{Yi:01, Demarque:04}. (a) Radius vs.\ mass, shown with
isochrones from 1~Gyr to 6~Gyr in steps of 1~Gyr (dotted lines) for a
fixed metallicity of [Fe/H] = +0.08. The best-fit isochrone for
3.85~Gyr is indicated with a solid line. The shaded area indicates the
uncertainty that comes from the 0.21~dex error in the measured
metallicity, at this particular age. (b) Effective temperature vs.\
mass.
\label{fig:yalemrt}}
\end{figure}

In Figure~\ref{fig:yalemrt}a we compare the two best determined
parameters of LV~Her ($M$ and $R$) against model isochrones from the
Yonsei-Yale series by \cite{Yi:01}, computed for the metallicity of
[Fe/H] $= +0.08$ determined in \S\thinspace\ref{sec:dimensions}. These
models include convective core overshooting
\citep[see][]{Demarque:04}, which has an effect for stars of this mass
and higher, and treats convection in the standard mixing length
approximation, with a mixing length parameter $\alpha_{\rm ML} =
1.7432$, calibrated against the Sun.  An isochrone corresponding to
3.85~Gyr (solid line) provides an excellent fit, and the high
precision of the measurements suggests an age uncertainty of no more
than 0.4~Gyr for this fixed metallicity.  However, the relatively
large error in the measured [Fe/H] adds significantly to the age
uncertainty. This is shown by the shaded area around the 3.85~Gyr
isochrone corresponding to $\sigma_{\rm [Fe/H]} = 0.21$~dex, which
contributes at least another 0.5~Gyr.  In the lower panel of the
figure the temperature is shown as a function of mass, with the same
models as in the top panel. Once again the isochrone of 3.85~Gyr
matches the observations very well, but in this case the lower
precision of $T_{\rm eff}$ provides a much weaker constraint on the
age, and gives rise to a larger uncertainty than the error that comes
from [Fe/H].

\begin{figure} 
\epsscale{1.0} 
\plotone{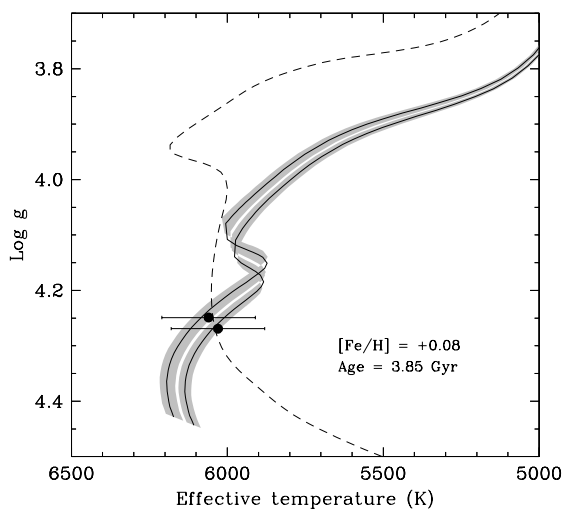}     
%
\figcaption[]{Surface gravity vs.\ effective temperature diagram for
LV~Her. The solid lines correspond to evolutionary tracks from the
Yonsei-Yale series \citep{Yi:01, Demarque:04} for the exact masses we
measure, and for the observed chemical composition. The effect of the
mass uncertainties in the location of these tracks is indicated by the
shaded areas. The dashed line represents the best-fit isochrone for
3.85~Gyr.\label{fig:yalelogg}}
\end{figure}

The $T_{\rm eff}$--$\log g$ diagram in Figure~\ref{fig:yalelogg}
implicitly compares all four measured quantities ($M$, $R$, $T_{\rm
eff}$, and [Fe/H]) against evolutionary tracks from the Yonsei-Yale
series. The models are computed for the observed metallicity of [Fe/H]
$= +0.08$, and for the precise masses we measure. The shaded area
indicates the uncertainty in the location of the tracks that comes
from the mass errors. The same 3.85~Gyr isochrone as before is shown
with a dashed line. The stars are seen to be in the middle of the
main-sequence band.

\begin{figure} 
\epsscale{1.0} 
\plotone{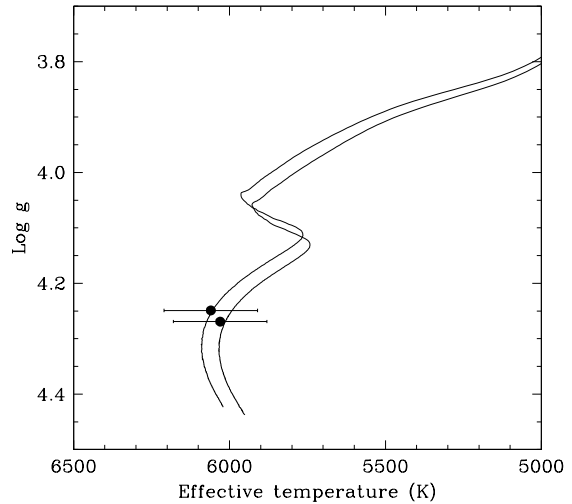}     
%
\figcaption[]{Surface gravity vs.\ effective temperature of LV~Her
compared with models by \cite{Claret:04}. The solid lines correspond
to evolutionary tracks for the measured masses, and for the observed
chemical composition. The age according to these models is
4.2~Gyr.\label{fig:claretlogg}}
\end{figure}

A similarly good fit to the properties of LV~Her is provided by the
stellar evolution models of \cite{Claret:04}, shown in
Figure~\ref{fig:claretlogg}, also for a metallicity constrained to the
measured value. These models are similar to the ones used before in
their treatment of convection ($\alpha_{\rm ML} = 1.68$), and also
include overshooting ($\alpha_{\rm ov} = 0.20$), but differ in the
details.  In this case, we find that the best-fit age is 4.2~Gyr,
about 9\% older than with the Yonsei-Yale models. The difference is of
the same order as the age uncertainty discussed above.

\subsection{Tidal theory}
\label{sec:tidal}

The high orbital eccentricity and our accurate measurement of the
projected rotational velocities of both components constitute useful
probes of tidal forces in the system and allow for an interesting
comparison with the predictions from theory. For our initial
comparisons we have used the radiative damping formalism of
\cite{Zahn:77} and \cite{Zahn:89}, often invoked for this type of
analysis, to calculate the critical times for LV~Her corresponding to
synchronization of the axial rotations, alignment of the rotational
and orbital axes, and circularization of the orbit. These are then
compared with the evolutionary age, which is 4.2~Gyr according to the
models by \citealt{Claret:04} used in this section. The procedure
follows closely that described by \cite{Claret:97}, and consists of
integrating the differential equations describing the changes in these
quantities along the evolutionary track for each star.

\begin{figure}[b]
\epsscale{1.15} 
\plotone{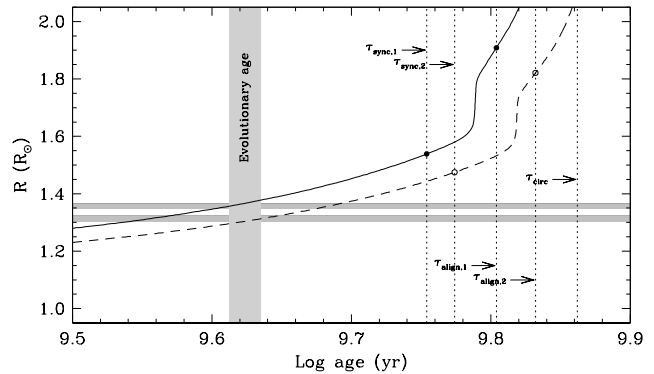}     
\figcaption[]{Radius as a function of age for LV~Her (solid line for
the primary, dashed for the secondary) from the stellar evolution
models by \cite{Claret:04}, for the measured masses and metallicity.
Measured radii and their uncertainties are represented by the
horizontal shaded bands. The critical times according to the theory by
\cite{Zahn:77, Zahn:89} are indicated with arrows for synchronization
and spin-orbit alignment of each component, as well as for
circularization of the orbit. The evolutionary age is also
shown.\label{fig:tidal1}}
\end{figure}

We find that rotational synchronization due to tidal forces is not
expected to occur for the primary until the system reaches an age of
$\tau_{\rm sync,1} = 5.7$~Gyr ($\log\tau_{\rm sync,1} = 9.754$), and
the secondary synchronization is predicted to happen shortly after
($\tau_{\rm sync,2} = 5.9$~Gyr, $\log\tau_{\rm sync,2} = 9.774$). This
appears to be in agreement with the fact that the measured $v \sin i$
values are slower than expected for pseudo-synchronization in the
eccentric orbit (Table~\ref{tab:dimensions}), assuming the spin axes
are parallel to the orbital axis (but see below). The time of orbital
circularization indicated by theory is 7.3~Gyr ($\log\tau_{\rm circ} =
9.862$), again consistent with the fact that the orbit is observed to
be highly eccentric. These times are displayed graphically in
Figure~\ref{fig:tidal1}, along with the evolution of the radius of
each star according to the evolutionary tracks.

A somewhat unexpected result is that the times of alignment predicted
by theory are also significantly longer than the current age of the
system: $\tau_{\rm align,1} = 6.4$~Gyr and $\tau_{\rm align,2} =
6.8$~Gyr (or $\log\tau_{\rm align,1} = 9.804$ and $\log\tau_{\rm
align,1} = 9.832$, in reference to Figure~\ref{fig:tidal1}). At face
value this indicates both spin axes may be inclined relative to the
axis of the orbit, which is the opposite of what is almost universally
assumed for binary systems. The long orbital period of LV~Her could
provide a plausible explanation, in principle, but in any case the
misalignment makes it somewhat problematic to interpret the apparent
lack of (pseudo-)synchronization inferred from the measured $v \sin
i$, since the projection factor can no longer be assumed to be known.

The calculations above involve a number of approximations implicit in
the equations we have used, as described, e.g., by \cite{Zahn:77} and
\cite{Hut:81}. In particular, they are linearized around the
equilibrium state, and are strictly valid only for relatively small
eccentricities and near-synchronous rotation, and for small relative
inclinations between the spin axes and the axis of the orbit. We have
also ignored changes in the semimajor axis that occur concurrently
with the evolution of other orbital elements. The condition on the
eccentricity is most certainly not met for LV~Her ($e \simeq 0.613$),
and it is unclear a priori to what extent this may affect the
conclusions above. This, and the potentially interesting situation
regarding the spin axes, have prompted us to explore the situation in
greater detail. We have directly integrated the more general
differential equations (valid also for high eccentricities) that
describe the evolution of the semimajor axis ($da/dt$), eccentricity
($de/dt$), angular rotational rates ($d\Omega_1/dt$, $d\Omega_2/dt$),
and inclination of the spin axes relative to the orbital axis
($d\Delta i/dt$ in our nomenclature) as given by \cite{Hut:81}, using
a fourth-order Runge-Kutta method.  Because they are coupled, these
equations must be integrated simultaneously, and as before the stellar
properties have been interpolated from the evolutionary tracks at each
time step. The turbulent dissipation timescale for stellar phases with
convective envelopes is taken to be $(MR^2/L)^{1/3}$, where $M$, $R$,
and $L$ are the mass, radius, and luminosity of the star. For phases
with radiative envelopes the timescales adopted follow closely those
in eq.(17) and eq.(18) by \cite{Claret:97}.

\begin{figure}[t]
\epsscale{1.16} 
\plotone{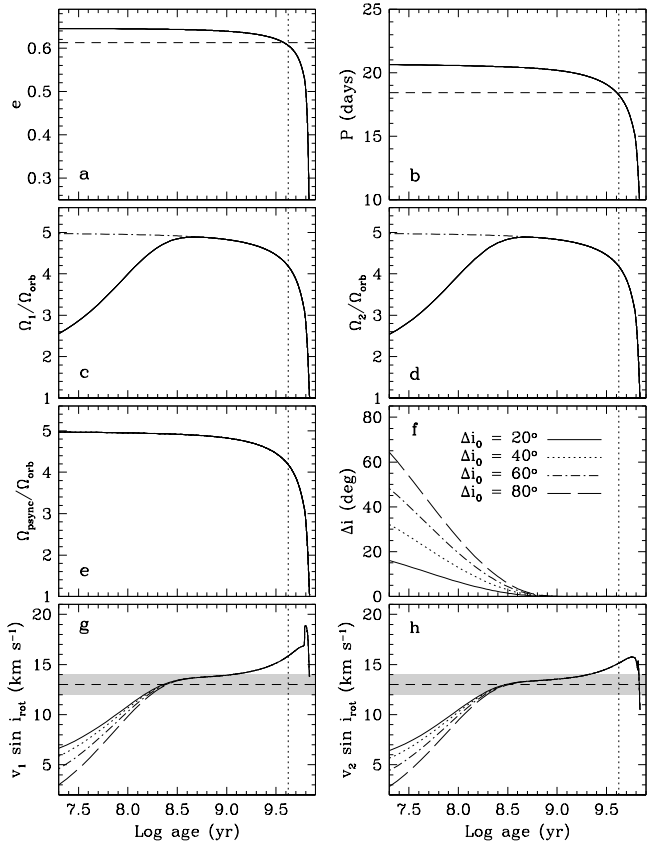}     
%
\figcaption[]{Observations of LV~Her compared against tidal theory
\citep{Hut:81}. (a) Eccentricity evolution. The dashed line represents
the measured value, and the vertical dotted line in this and the other
panels marks the current evolutionary age of 4.2~Gyr, according to the
models by \cite{Claret:04}. (b) Evolution of the orbital period, with
the current value represented with a dashed line. (c) Rotation rate of
the primary, normalized to the orbital rate. The dot-dashed line
represents the evolution of the pseudo-synchronous rate shown in panel
(e). (d) Same as (c), for the secondary. (e) Evolution of the
pseudo-synchronous rotation rate. (f) Evolution of the relative angle
$\Delta i$ between the spin axis and the orbital axis, for different
initial values. (g) Evolution of the projected rotational velocity of
the primary, for the same four trial values of $\Delta i$. The dashed
line represents the measured value, its uncertainty indicated by the
shaded area. (h) Same as (g), for the secondary.\label{fig:tidal2}}
\end{figure}

The initial conditions are of course not known, so we explored a range
of values that give predictions matching the observed quantities at
the current age of the system. This is shown in
Figure~\ref{fig:tidal2}a,b for the eccentricity and orbital
period. The initial values are $e_0 = 0.66$ and $P_0 = 21.8$~days at
the starting age of $\log \tau_0 = 7.0$. According to these
calculations we expect the orbit to circularize at an age $\log
\tau_{\rm circ} \simeq 9.85$ ($\tau_{\rm circ} \simeq 7.1$~Gyr), which
is in fairly good agreement with our previous estimate despite the
concerns expressed above.

Also shown in the figure is the evolution of the angular rotation rate
$\Omega = 2\pi/P_{\rm rot}$ for each star, which we normalize for
convenience to the orbital rate $\Omega_{\rm orb}= 2\pi/P_{\rm orb}$
(Figure~\ref{fig:tidal2}c,d; solid lines).  For LV~Her we have no
observational constraints on $\Omega_1$ and $\Omega_2$, so the initial
values are arbitrary and chosen to be the same. In principle they
could be constrained by directly measuring the rotation periods of the
stars due, e.g., to rotational modulation by spots, although in
practice this is difficult for a spatially unresolved binary. More
importantly, these stars may be too hot to be affected significantly
by spots.  The ratio between the pseudo-synchronous velocity and the
mean orbital motion is shown in Figure~\ref{fig:tidal2}e, and this
curve is repeated in Figure~\ref{fig:tidal2}c,d (dot-dashed line). As
seen from the convergence of the dot-dashed and solid curves, the
stars are predicted to reach pseudo-synchronization roughly at $\log
\tau_{\rm psync} \simeq 8.75$ ($\tau_{\rm psync} \simeq 0.6$~Gyr),
which is much younger than the evolutionary age. This is completely at
odds with our previous estimate, showing the limitations of that
approach for a system like LV~Her.

We also have no direct constraint on the relative angle $\Delta i$
between the spin axes and the orbital axis. Furthermore, the relation
between $\Delta i$ and the orbital and rotational inclinations $i_{\rm
orb}$ and $i_{\rm rot}$, both measured relative to the plane of the
sky, is given by
\begin{equation}
\cos \Delta i = \cos i_{\rm orb} \cos i_{\rm rot} + \sin i_{\rm orb} \sin i_{\rm rot} \cos \lambda~,
\end{equation}
which involves an unknown angle $\lambda$ between the sky-projected
angular momentum vectors of the orbit and the stellar spin.

The spectroscopically measured projected rotational velocities of the
stars, which we refer to more properly now as $v \sin i_{\rm rot}$, do
provide some indirect constraint on a combination of the theoretically
predictable quantities, but this still involves the unknown angle
$\lambda$.  Because $i_{\rm orb}$ is very nearly 90\arcdeg\ for LV~Her
(see Table~\ref{tab:dimensions}), we may make the approximation $\cos
\Delta i \approx \sin i_{\rm rot} \cos \lambda$. We then have
\begin{equation}
\label{eq:tidal}
v_{1,2} \sin i_{\rm rot} \approx {2\pi\over P_{\rm orb}} {\Omega_{1,2}\over \Omega_{\rm orb}} {\cos\Delta i\over \cos \lambda} R~.
\end{equation}
In this equation all quantities on the right-hand side are either
known from stellar evolution calculations ($R$), or can be computed
from the solution of the differential equations, except for
$\cos\lambda$, which depends on the observer's viewpoint.  For the
sake of illustration, we ignore this term in the following (or
consider $\lambda$ to be small), so that $\cos \Delta i \approx \sin
i_{\rm rot}$.  In Figure~\ref{fig:tidal2}f we show the evolution of
$\Delta i$ for four different initial values (20\arcdeg, 40\arcdeg,
60\arcdeg, 80\arcdeg). Curves for the primary and secondary are nearly
indistinguishable, so we show only those for the primary. Once again
the conclusion is very different from the one we reached with the
equations by Zahn: alignment of the spin axes with the orbital axis is
achieved much earlier than the age of the system, at $\log \tau_{\rm
align} \simeq 8.9$ (or $\tau_{\rm align} \simeq 0.8$~Gyr).  The
corresponding $v \sin i_{\rm rot}$ curves for the primary and
secondary, calculated using eq.(\ref{eq:tidal}), are shown in the
bottom panels and are compared with the measured rotations (dashed
line and shaded uncertainty region). The general trend is for the
stars in LV~Her to be spun up by tidal forces, but the calculations
disagree with the observations at the current age of the system,
predicting the measured rotations should be 16~\kms\ for the primary
and 15.5~\kms\ for the secondary (see also
Table~\ref{tab:dimensions}).\footnote{Although the rotation rates
$\Omega_1$ and $\Omega_2$ have been chosen arbitrarily here, this does
not affect the predicted rotational velocities at the current age
because pseudo-synchronization occurs much earlier, independently of
the starting values of $\Omega_1$ and $\Omega_2$.} These are nominally
outside of the range allowed by the observational uncertainties. We
consider it very unlikely that our measurements are off by as much as
3~\kms\ for the primary and 2.5~\kms\ for the secondary, since our
formal uncertainty of $\pm$1~\kms\ is already conservative.  The
alternative would be a deficiency in the calculations, related to
remaining approximations in the differential equations for this very
challenging problem. A combination of both effects is also
possible. An even more general treatment of the dynamical problem of
tidal friction in binary systems with deformable components has been
developed by \cite{Alexander:73}, but the application is considerably
more complex and is beyond the scope of the present work.

We note, finally, that an empirical test of the prediction of
alignment is available in eclipsing binaries by measuring the
Rossiter-McLaughlin effect \citep[e.g.,][]{Albrecht:07,
Albrecht:08}. This provides a direct measure of the angle $\lambda$
between the sky projections of the spin axes and the axis of the
orbit.

\section{Discussion and conclusions}
\label{sec:discussion}

As a result of our intensive spectroscopic and photometric monitoring,
the absolute masses and radii of LV~Her are now determined to 0.9\% or
better, and are among the best available for any eclipsing binary.  In
addition we have established the effective temperatures, as well as
the overall metallicity. The latter quantity has not been determined
for many eclipsing binaries, and is important because it reduces the
degrees of freedom in the comparison with theory. Unfortunately the
precision of [m/H] (and $T_{\rm eff}$ to some extent) is limited in
this case by poor knowledge of interstellar extinction in the
direction of the star. This could perhaps be remedied with additional
absolute photometric observations. Nevertheless, the combination of
$M$, $R$, $T_{\rm eff}$, and [m/H] provides unusually strong
contraints on stellar evolution theory. Comparison with current models
yields an excellent fit to the observations, and indicates the stars
are approximately half way through their main-sequence phase, at an
age between 3.8 and 4.2~Gyr, depending on the model.

Eclipsing binaries with periods as long as that of LV~Her are
relatively rare and difficult to find, except perhaps in the course of
automated variability surveys such as those designed to search for
transiting planets, which often operate almost continuously for weeks
or months at a time. The very high eccentricity of the system ($e
\simeq 0.613$) makes it the most extreme case among those with
well-determined properties \citep{Torres:09}. Although apsidal motion
is expected, the apsidal period is predicted to be very long ($U \sim
58,\!000$~yr) on account of the wide separation between the stars; it
may take decades for the effect to be detected and measured
accurately.

Tidal theory is consistent with the observations in predicting that
the orbit should not yet have been circularized by tidal
forces. Direct integration of the coupled differential equations for
this problem according to \cite{Hut:81} indicates that the spin axes
have already been forced into alignment with the orbital axis early on
in the evolution of this system. The spins of both stars are also
expected to be pseudo-synchronized with the orbital motion, although
this is not quite in agreement with the measured rotations, and may
indicate either large errors of measurement (which we believe are
unlikely) or shortcomings in theory, or some combination of both. The
case of LV~Her highlights the importance of bearing in mind the
assumptions under which the equations for tidal evolution have been
derived.  Because of their ease of use and simpler interpretation, the
linearized equations by \cite{Zahn:77} \citep[see also][]{Hut:81,
Claret:97} have occasionally been applied to binary systems that may
violate those assumptions to some degree (namely, the requirement of
small eccentricities and spin-orbit inclinations, and near-synchronous
rotations). Exactly when and how the theory breaks down is difficult
to predict. With its high eccentricity, LV~Her is one such case, as we
have demonstrated here by comparing the critical times from the
simpler formulation with the results from the more general equations
given by \cite{Hut:81}.

\acknowledgements

We are grateful to P.\ Berlind, M.\ Calkins, R.\ J.\ Davis, E.\
Horine, D.\ W.\ Latham, J.\ Peters, and J.\ Zajac for their help in
gathering the spectroscopic observations of LV~Her used in this work,
and to R.\ J.\ Davis for maintaining the echelle database at CfA.  GT
acknowledges partial support for this research through NSF grant
AST-0708229.  CHSL thanks Bill Neely who operates and maintains the
NFO for our Consortium, and who handles preliminary processing and
storage of the images.  He also wishes to thank summer 2009 Arkansas
REU student Chris Gong for her preliminary analysis of the URSA
data. This research has made use of the SIMBAD database and the VizieR
catalogue access tool, both operated at CDS, Strasbourg, France, of
NASA's Astrophysics Data System Abstract Service, and of data products
from the Two Micron All Sky Survey (2MASS), which is a joint project
of the University of Massachusetts and the Infrared Processing and
Analysis Center/California Institute of Technology, funded by NASA and
the NSF.


\begin{thebibliography}{}

\bibitem[Albrecht(2008)]{Albrecht:08} Albrecht, S. 2008, Stars and
planets at high spatial and spectral resolution, PhD Thesis, Leiden
University, The Netherlands

\bibitem[Albrecht et al.(2007)]{Albrecht:07}
 Albrecht, S., Reffert, S., Snellen, I., Quirrenbach, A., \& Mitchell,
 D.\ S. 2007, \aap, 474, 565

\bibitem[Alexander(1973)]{Alexander:73}
 Alexander, M.\ E. 1973, \apss, 23, 459

\bibitem[Br\'at et al.(2007)]{Brat:07}
 Br\'at, L., Zejda, M., \& Svoboda, P. 2007, B.R.N.O.\ Contrib., 34

\bibitem[Casagrande et al.(2006)]{Casagrande:06}
 Casagrande, L., Portinari, L., \& Flynn, C. 2006, \mnras, 373, 13

\bibitem[Claret \& Cunha(1997)]{Claret:97}
 Claret, A., \& Cunha, N.\ C.\ S. 1997, \aap, 318, 187

\bibitem[Claret(1998)]{Claret:98}
 Claret, A. 1998, \aaps, 131, 395

\bibitem[Claret(2000)]{Claret:00}
 Claret, A. 2000, \aap, 383, 1081

\bibitem[Claret(2004)]{Claret:04}
 Claret, A. 2004, \aap, 424, 919

\bibitem[Claret(2008)]{Claret:08}
 Claret, A. 2008, \aap, 482, 259

\bibitem[Demarque et al.(2004)]{Demarque:04}
 Demarque, P., Woo, J.-H., Kim, Y.-C., \& Yi, S.\ K. 2004, \apjs, 155,
667

\bibitem[Diethelm(2008)]{Diethelm:08}
 Diethelm, R. 2008, IBVS, No.\ 5837

\bibitem[Droege et al.(2007)]{Droege:07}
 Droege, T. F., Richmond, M.\ W., \& Sallman, M. 2006, \pasp, 118,
 1666

\bibitem[Etzel(1981)]{Etzel:81}
 Etzel, P.\ B. 1981, Photometric and Spectroscopic Binary Systems
(Dordrecht: Reidel), 65

\bibitem[Flower(1996)]{Flower:96}
 Flower, P.\ J. 1996, \apj, 469, 355

\bibitem[Gonz\'alez Hern\'andez \& Bonifacio(2009)]{Gonzalez:09}
 Gonz\'alez Hern\'andez, J.\ I., \& Bonifacio, P. 2009, \aap, 497, 497

\bibitem[Grauer et al.(2008)]{Grauer:08}
 Grauer, A.\ D., Neely, A.\ W., \& Lacy, C.\ H.\ S. 2008, \pasp, 120,
 992

\bibitem[Hilditch \& Hill(1975)]{Hilditch:75}
 Hilditch, R.\ W. \& Hill, G. 1975, \mnras, 79, 101

\bibitem[Hoffmeister(1935)]{Hoffmeister:35}
 Hoffmeister, C. 1935, \an, 255, 401

\bibitem[H\o g et al.(2000)]{Hog:00}
 H\o g, E., Fabricius, C., Makarov, V.\ V., Urban, S., Corbin, T.,
 Wycoff, G., Bastian, U., Schwekendiek, P., \& Wicenec, A. 2000, \aap,
 355, L27

\bibitem[H\"ubscher(2005)]{Hubscher:05}
 H\"ubscher, J. 2005, IBVS, No.\ 5643

\bibitem[H\"ubscher et al.(2008)]{Hubscher:08}
 H\"ubscher, J., Steinbach, H.-M., \& Walter, F. 2008, IBVS, No.\ 5830

\bibitem[Hut(1981)]{Hut:81}
 Hut P. 1981, \aap, 99, 126

\bibitem[Lacy(2002)]{Lacy:02} 
 Lacy, C.\ H.\ S. 2002, IBVS, No.\ 5357

\bibitem[Lacy(2003)]{Lacy:03} 
 Lacy, C.\ H.\ S. 2003, IBVS, No.\ 5487

\bibitem[Lacy(2004)]{Lacy:04} 
 Lacy, C.\ H.\ S. 2004, IBVS, No.\ 5577

\bibitem[Lacy(2006)]{Lacy:06} 
 Lacy, C.\ H.\ S. 2006, IBVS, No.\ 5670

\bibitem[Lacy(2007)]{Lacy:07} 
 Lacy, C.\ H.\ S. 2007, IBVS, No.\ 5764

\bibitem[Lacy et al.(2004a)]{Lacy:04a}
 Lacy, C.\ H.\ S., Claret, A., \& Sabby, J.\ A. 2004a, \aj, 128, 1840

\bibitem[Lacy et al.(2004b)]{Lacy:04b}
 Lacy, C.\ H.\ S., Claret, A., Sabby, J.\ A., Hood, B., \& Secosan,
F. 2004b, \aj, 128, 3005

\bibitem[Lacy et al.(2001)]{Lacy:01}
 Lacy, C.\ H.\ S., Hood, B., \& Straughn, A. 2001, IBVS, No.\ 5067

\bibitem[Lacy et al.(2002)]{Lacyetal:02}
 Lacy, C.\ H.\ S., Straughn, A., \& Denger, F. 2002, IBVS, No.\ 5251

\bibitem[Lacy et al.(2008)]{Lacy:08}
 Lacy, C.\ H.\ S., Torres, G., \& Claret, A. 2008, \aj, 135, 1757

\bibitem[Lacy et al.(2006)]{Lacyetal:06}
 Lacy, C.\ H.\ S., Torres, G., Claret, A., \& Menke, J.\ L. 2006, \aj,
131, 2664

\bibitem[Lacy et al.(2005)]{Lacy:05}
 Lacy, C.\ H.\ S., Torres, G., Claret, A., \& Vaz, L.\ P.\ R. 2005,
\aj, 130, 2838

\bibitem[Latham(1985)]{Latham:85}
 Latham, D.\ W. 1985, in IAU Coll.\ 88, Stellar Radial Velocities,
 eds.\ A.\ G.\ D.\ Philip \& D.\ W.\ Latham (Schenectady: L.\ Davis),
 21

\bibitem[Latham(1992)]{Latham:92}
 Latham, D.\ W. 1992, in IAU Coll.\ 135, Complementary Approaches to
Double and Multiple Star Research, ASP Conf.\ Ser.\ 32, eds.\ H.\ A.\
McAlister \& W.\ I.\ Hartkopf (San Francisco: ASP), 110

\bibitem[Latham et al.(1996)]{Latham:96}
 Latham, D.\ W., Nordstr\"om, B., Andersen, J., Torres, G., Stefanik,
R.\ P., Thaller, M., \& Bester, M. 1996, \aap, 314, 864

\bibitem[Latham et al.(2002)]{Latham:02}
 Latham, D.\ W., Stefanik, R.\ P., Torres, G., Davis, R.\ J., Mazeh,
 T., Carney, B.\ W., Laird, J.\ B., \& Morse, J.\ A. 2002, \aj, 124,
 1144

\bibitem[Perry \& Johnson(1982)]{Perry:82}
 Perry, C.\ L., \& Johnson, L. 1982, \apjs, 50, 451

\bibitem[Perryman et al.(1997)]{Perryman:97}
 Perryman, M.\ A.\ C., et al. 1997, The \hip\ and Tycho Catalogues
 (ESA SP-1200; Noordwjik: ESA)

\bibitem[Popper(1996)]{Popper:96}
 Popper, D.\ M. 1996, \apjs, 106, 133

\bibitem[Popper(1980)]{Popper:80}
 Popper, D.\ M. 1980, \araa, 18, 115

\bibitem[Popper \& Etzel(1981)]{Popper:81}
 Popper, D.\ M., \& Etzel, P.\ B. 1981, \aj, 86, 102
	
\bibitem[Ram\'\i rez \& Mel\'endez(2005)]{Ramirez:05}
 Ram\'\i rez, I., \& Mel\'endez, J. 2005, \apj, 626, 465

\bibitem[Schlegel et al.(1998)]{Schlegel:98}
 Schlegel, D.\ J., Finkbeiner, D.\ P., \& Davis, M. 1998, \apj, 500,
525

\bibitem[Selman(2004)]{Selman:04}
 Selman F.\ J. 2004, Optimizing Scientific Return for Astronomy
through Information Technologies, Proc.\ SPIE, 5493, 453

\bibitem[Southworth et al.(2007)]{Southworth:07}
 Southworth, J., Bruntt, H., \& Buzasi, D.\ L. 2007, \aap, 467, 1215

\bibitem[Torres(2000)]{Torres:00}
 Torres, G. 2000, IBVS No.\ 4971

\bibitem[Torres et al.(2009)]{Torres:09}
 Torres, G., Andersen, J., \& Gim\'enez, A. 2009, \aapr, in press

\bibitem[Torres et al.(2000a)]{Torresetal:00a}
 Torres, G., Andersen, J., Nordstr\"om, B., \& Latham, D.\ W. 2000a,
\aj, 119, 1942

\bibitem[Torres et al.(2000b)]{Torresetal:00b}
 Torres, G., Lacy, C.\ H.\ S., Claret, A., \& Sabby, J.\ A. 2000b, \aj,
 120, 3226

\bibitem[Torres et al.(2001)]{Torres:01}
 Torres, G., Lacy, C.\ H.\ S., Guilbault, P.\ R., Diethelm, R.,
Baldwin, M.\ E., \& Lubcke, G.\ C. 2001, IBVS No.\ 5201

\bibitem[Torres et al.(2002)]{Torres:02}
 Torres, G., Neuh\"auser, R., \& Guenther, E.\ W. 2002, \aj, 123, 1701

\bibitem[Torres et al.(1997)]{Torres:97}
 Torres, G., Stefanik, R.\ P., Andersen, J., Nordstr\"om, B., Latham,
D.\ W., \& Clausen. J.\ V. 1997, \aj, 114, 2764

\bibitem[van Hamme(1993)]{vanHamme:93}
 van Hamme, W. 1993, \aj, 106, 2096

\bibitem[Yi et al.(2001)]{Yi:01}
 Yi, S.\ K., Demarque, P., Kim, Y.-C., Lee, Y.-W., Ree, C.\ H.,
Lejeune, T., \& Barnes, S. 2001, \apjs, 136, 417

\bibitem[Zahn(1977)]{Zahn:77}
 Zahn, J.-P. 1977, \aap, 57, 383

\bibitem[Zahn(1989)]{Zahn:89}
 Zahn, J.-P. 1989, \aap, 220, 112

\bibitem[Zessewitsch(1944)]{Zessewitsch:44}
 Zessewitsch, V.\ P. 1944, Astron.\ Circ., 36, 5

\bibitem[Zessewitsch(1954)]{Zessewitsch:54}
 Zessewitsch, V.\ P. 1954, Odessa Report, 4, 113

\bibitem[Zucker \& Mazeh(1994)]{Zucker:94} Zucker, S., \& Mazeh,
T. 1994, \apj, 420, 806

\end{thebibliography}
\end{document}